\documentclass[tighten,times,twocolumn]{aastex62}
\pdfoutput=1
\shorttitle{Spectroscopic Confirmation of a Protocluster at $z=3.37$}
\shortauthors{McConachie {\em et al.}}

\newcommand{\eg}{{\rm e.g.,}}
\newcommand{\ie}{{\rm i.e.,}}

\newcommand{\Mo}{\ensuremath{{\rm M}_\odot}}

\newcommand{\Mstar}{\ensuremath{M_{\star}}}

\newcommand{\amin}{\ensuremath{\arcmin}}

\newcommand{\lUMGA}{{\rm UMG}~ID~``COS-DR3-\ensuremath{179370}''}
\newcommand{\lUMGB}{{\rm UMG}~ID~``COS-DR3-\ensuremath{160748}''}
\newcommand{\UMGA}{{\rm UMG}~\ensuremath{179370}}
\newcommand{\UMGB}{{\rm UMG}~\ensuremath{160748}}
\newcommand{\A}{\ensuremath{179370}}
\newcommand{\B}{\ensuremath{160748}}

\newcommand{\Hbeta}{$\rm{H}\beta$}
\newcommand{\Halpha}{$\rm{H}\alpha$}

\newcommand{\OIII}{[O{\rm \scriptsize III}]$\lambda\lambda4959,5007$}

\newcommand{\logM}{$\log(M_{\star}/{\rm M}_\odot)$}
\newcommand{\surveyname}{MAGAZ3NE}
\newcommand{\protoclusterA}{MAGAZ3NE J095924+022537} 
\newcommand{\protoclusterB}{MAGAZ3NE J100028+023349}
\newcommand{\pA}{MAGAZ3NE J0959} 
\newcommand{\pB}{MAGAZ3NE J1000}
\newcommand{\pAB}{MAGAZ3NE J0959/J1000}

\newcommand\aastexcls{2}
\newcommand\othercls{3}

\newcommand\papercls{\aastexcls}

\usepackage{apjfonts}

\usepackage{ifthen}
\usepackage{natbib}
\usepackage{amssymb, amsmath}
\usepackage{appendix}
\usepackage{etoolbox}
\usepackage[T1]{fontenc}
\usepackage{paralist}

\if\papercls \othercls

\else

\fi

\providecommand{\adsurl}[1]{\href{#1}{ADS}}

\makeatletter
\patchcmd{\NAT@citex}
  {\@citea\NAT@hyper@{%
     \NAT@nmfmt{\NAT@nm}%
     \hyper@natlinkbreak{\NAT@aysep\NAT@spacechar}{\@citeb\@extra@b@citeb}%
     \NAT@date}}
  {\@citea\NAT@nmfmt{\NAT@nm}%
   \NAT@aysep\NAT@spacechar\NAT@hyper@{\NAT@date}}{}{}

\patchcmd{\NAT@citex}
  {\@citea\NAT@hyper@{%
     \NAT@nmfmt{\NAT@nm}%
     \hyper@natlinkbreak{\NAT@spacechar\NAT@@open\if*#1*\else#1\NAT@spacechar\fi}%
       {\@citeb\@extra@b@citeb}%
     \NAT@date}}
  {\@citea\NAT@nmfmt{\NAT@nm}%
   \NAT@spacechar\NAT@@open\if*#1*\else#1\NAT@spacechar\fi\NAT@hyper@{\NAT@date}}
  {}{}
\makeatother

\makeatletter
\DeclareRobustCommand{\lowcase}[1]{\@lowcase#1\@nil}
\def\@lowcase#1\@nil{\if\relax#1\relax\else\MakeLowercase{#1}\fi}
\makeatother

\DeclareSymbolFont{UPM}{U}{eur}{m}{n}
\DeclareMathSymbol{\umu}{0}{UPM}{"16}
\let\oldumu=\umu
\renewcommand\umu{\ifmmode\oldumu\else\math{\oldumu}\fi}

\if\papercls \othercls

\else

\fi

\let\oldsim=\sim
\renewcommand\sim{\ifmmode\oldsim\else\math{\oldsim}\fi}
\let\oldpm=\pm
\renewcommand\pm{\ifmmode\oldpm\else\math{\oldpm}\fi}
\newcommand\by{\ifmmode\times\else\math{\times}\fi}

\newbox{\wdbox}
\renewcommand\c{\setbox\wdbox=\hbox{,}\hspace{\wd\wdbox}}
\renewcommand\i{\setbox\wdbox=\hbox{i}\hspace{\wd\wdbox}}

\newcount\timect
\newcount\hourct
\newcount\minct
\newcommand\now{\timect=\time \divide\timect by 60
         \hourct=\timect \multiply\hourct by 60
         \minct=\time \advance\minct by -\hourct
         \number\timect:\ifnum \minct < 10 0\fi\number\minct}

\catcode`@=11

\newcommand\comment[1]{}

\newcommand\commenton{\catcode`\%=14}

\renewcommand\math[1]{$#1$}
\newcommand\mathshifton{\catcode`\$=3}

\let\atab=&
\newcommand\atabon{\catcode`\&=4}

\let\oldmsp=\sp
\let\oldmsb=\sb
\def\sp#1{\ifmmode
           \oldmsp{#1}%
         \else\strut\raise.85ex\hbox{\scriptsize #1}\fi}
\def\sb#1{\ifmmode
           \oldmsb{#1}%
         \else\strut\raise-.54ex\hbox{\scriptsize #1}\fi}
\newbox\@sp
\newbox\@sb
\def\sbp#1#2{\ifmmode%
           \oldmsb{#1}\oldmsp{#2}%
         \else
           \setbox\@sb=\hbox{\sb{#1}}%
           \setbox\@sp=\hbox{\sp{#2}}%
           \rlap{\copy\@sb}\copy\@sp
           \ifdim \wd\@sb >\wd\@sp
             \hskip -\wd\@sp \hskip \wd\@sb
           \fi
        \fi}
\def\msp#1{\ifmmode
           \oldmsp{#1}
         \else \math{\oldmsp{#1}}\fi}
\def\msb#1{\ifmmode
           \oldmsb{#1}
         \else \math{\oldmsb{#1}}\fi}

\def\supon{\catcode`\^=7}

\def\subon{\catcode`\_=8}

\def\supsubon{\supon \subon}

\newcommand\actcharon{\catcode`\~=13}

\newcommand\paramon{\catcode`\#=6}

\comment{And now to turn us totally on and off...}

\newcommand\reservedcharson{ \commenton  \mathshifton  \atabon  \supsubon 
                             \actcharon  \paramon}

\catcode`@=12
\reservedcharson

\if\papercls \aastexcls

\else

\fi

\newcommand\chisq{\ifmmode{\chi\sp{2}}\else\math{\chi\sp{2}}\fi}
\newcommand\redchisq{\ifmmode{ \chi\sp{2}\sb{\rm red}}
                    \else\math{\chi\sp{2}\sb{\rm red}}\fi}
\newcommand\Teq{\ifmmode{T\sb{\rm eq}}\else$T$\sb{eq}\fi}
\newcommand\mjup{\ifmmode{M\sb{\rm Jup}}\else$M$\sb{Jup}\fi}
\newcommand\rjup{\ifmmode{R\sb{\rm Jup}}\else$R$\sb{Jup}\fi}
\newcommand\msun{\ifmmode{M\sb{\odot}}\else$M\sb{\odot}$\fi}
\newcommand\rsun{\ifmmode{R\sb{\odot}}\else$R\sb{\odot}$\fi}
\newcommand\mearth{\ifmmode{M\sb{\oplus}}\else$M\sb{\oplus}$\fi}
\newcommand\rearth{\ifmmode{R\sb{\oplus}}\else$R\sb{\oplus}$\fi}

\begin{document}

\title{Spectroscopic Confirmation of a Protocluster at $z=3.37$ with a High Fraction of Quiescent Galaxies}

\author[0000-0002-2446-8770]{Ian McConachie}
\affiliation{Department of Physics and Astronomy, University of California, Riverside, 900 University Avenue, Riverside, CA 92521, USA}

\author[0000-0002-6572-7089]{Gillian Wilson}
\affiliation{Department of Physics and Astronomy, University of California, Riverside, 900 University Avenue, Riverside, CA 92521, USA}

\author[0000-0001-6003-0541]{Ben Forrest}
\affiliation{Department of Physics and Astronomy, University of California, Riverside, 900 University Avenue, Riverside, CA 92521, USA}

\author[0000-0002-7248-1566]{Z. Cemile Marsan}
\affiliation{Department of Physics and Astronomy, York University, 4700, Keele Street, Toronto, ON MJ3 1P3, Canada}

\author[0000-0002-9330-9108]{Adam Muzzin}
\affiliation{Department of Physics and Astronomy, York University, 4700, Keele Street, Toronto, ON MJ3 1P3, Canada}

\author[0000-0003-1371-6019]{M. C. Cooper}
\affiliation{Center for Cosmology, Department of Physics and Astronomy, University of California, Irvine, 4129 Frederick Reines Hall, Irvine, CA, USA}

\author{Marianna Annunziatella}
\affiliation{Department of Physics and Astronomy, Tufts University, 574 Boston Avenue Suites 304, Medford, MA 02155, USA}
\affiliation{Centro de Astrobiolog\'{i}a (CSIC-INTA), Ctra de Torrej\'{o}n a Ajalvir, km 4, E-28850 Torrej\'{o}n de Ardoz, Madrid, Spain}

\author[0000-0001-9002-3502]{Danilo Marchesini}
\affiliation{Department of Physics and Astronomy, Tufts University, 574 Boston Avenue Suites 304, Medford, MA 02155, USA}

\author[0000-0001-6251-3125]{Jeffrey C. C. Chan}
\affiliation{Department of Physics and Astronomy, University of California, Riverside, 900 University Avenue, Riverside, CA 92521, USA}

\author{Percy Gomez}
\affiliation{W.M. Keck Observatory, 65-1120 Mamalahoa Hwy., Kamuela, HI 96743, USA}

\author[0000-0003-3595-7147]{Mohamed H. Abdullah}
\affiliation{Department of Physics and Astronomy, University of California, Riverside, 900 University Avenue, Riverside, CA 92521, USA}
\affiliation{Department of Astronomy, National Research Institute of Astronomy and Geophysics, Helwan, 11421, Egypt}

\author[0000-0003-3959-2595]{Paolo Saracco}
\affiliation{INAF - Osservatorio Astronomico di Brera, via Brera 28, 20121 Milano, Italy}

\author[0000-0002-7356-0629]{Julie Nantais}
\affiliation{Departanento de Ciencias F\'{i}sicas, Universidad Andr\'{e}s Bello, Fern\'{a}ndez Concha 700, Las Condes, Santiago, Chile}

\correspondingauthor{Ian McConachie}
\email{ian.mcconachie@email.ucr.edu}

\revised{August 27, 2021}
 \accepted{September 28, 2021}
\submitjournal{ApJ}

\begin{abstract}
  We report the discovery of \protoclusterA, a spectroscopically-confirmed protocluster at $z = 3.3665^{+0.0009}_{-0.0012}$ around a spectroscopically-confirmed $UVJ$-quiescent ultra-massive galaxy (UMG; $\Mstar=2.34^{+0.23}_{-0.34}\times10^{11}~\Mo$) in the COSMOS UltraVISTA field. We present a total of 38 protocluster members (14 spectroscopic and 24 photometric), including the UMG. Notably, and in marked contrast to protoclusters previously reported at this epoch which have been found to contain predominantly star-forming members, we measure an elevated fraction of quiescent galaxies relative to the coeval field ($73.3^{+26.7}_{-16.9}\%$ versus $11.6^{+7.1}_{-4.9}\%$ for galaxies with stellar mass $\Mstar \geq 10^{11} \Mo$). This high quenched fraction provides a striking and important counterexample to the seeming ubiquitousness of star-forming galaxies in protoclusters at $z>2$ and suggests, rather, that protoclusters exist in a diversity of evolutionary states in the early Universe. We discuss the possibility that we might be observing either ``early mass quenching'' or non-classical ``environmental quenching.'' We also present the discovery of \protoclusterB, a second spectroscopically-confirmed protocluster, at a very similar redshift of $z = 3.3801^{+0.0213}_{-0.0281}$. We present a total of 20 protocluster members, 12 of which are photometric and 8 spectroscopic including a post-starburst UMG ($\Mstar=2.95^{+0.21}_{-0.20}\times10^{11}~\Mo$). Protoclusters \pA\ and \pB\ are separated by 18 arcminutes on the sky (35 comoving Mpc), in good agreement with predictions from simulations for the size of ``Coma''-type cluster progenitors at this epoch. It is highly likely that the two UMGs are the progenitors of Brightest Cluster Galaxies (BCGs) seen in massive virialized clusters at lower redshift.
\end{abstract}




\section{Introduction}
\label{introduction}

In the local Universe, massive clusters (total mass  $\geq~10^{15}~\Mo$) extend over only a few Mpc (\eg\ \citealt{abdullah-20a}). However, numerical simulations have shown that the progenitors of these present-day clusters are very much more extended \citep{angulo-12, chiang-13, muldrew-15, overzier-16}. For example, using the Millennium Simulation (\citealt{springel-05}), \citet{muldrew-15} found that, at $3 < z < 4$,  $90\%$ of the stellar mass of a protocluster with total mass $M_{z=0} = 10^{15.4}~h^{-1}\Mo$ typically extends over 65 comoving Mpc, corresponding to 14.5 physical Mpc or 28.5 arcmin on the sky. In order to comprehensively study massive protoclusters at high redshift, observations spanning tens of arcmins on the sky are required.

Closely related to the question of how protoclusters form in the early Universe is how galaxies evolve within them. It has long been established that in the local Universe there exists a strong dependence of galaxy properties on environment \ie\ denser environments result in higher fractions of early-type galaxies (\citealt{oemler-74, dressler-80, binggeli-87, goto-03, fasano-12, fasano-15}) and enhanced quenched fractions (\citealt{gomez-03, kauffmann-04, peng-10, wetzel-12, wang-18,roberts-19, Li2020}). This is because galaxies in dense environments have been subject to ``extra'' external processes (environmental quenching) such as ram pressure stripping (\citealt{gunn-72}), galaxy-galaxy interactions (\citealt{farouki-81}), harassment (\citealt{moore-96}),  and strangulation (\citealt{larson-80}). These processes are in addition to ``regular'' internal processes
(mass quenching), such as active galactic nuclei (AGN; \citealt{fabian-12}) and stellar feedback (\citealt{hopkins-14a}). One of the most interesting questions in galaxy evolution is at what epoch environmental quenching first begins to take effect.


The most robust indicator of the presence of environmental quenching is an increase in the  fraction of quiescent galaxies relative to that measured in the coeval field. Higher quiescent fractions have been observed out to $z\sim2$, both directly from analysis of spectroscopically-confirmed clusters and groups  (\citealt{muzzin-12, quadri-12, newman-14, balogh-16, cooke-16, lee-brown-17, nantais-16, nantais-17,  lemaux-19, pintos-castro-19, strazzullo-19, zavala-19, vanderBurg-20}), and indirectly from statistical analysis of photometric overdensities (\citealt{cooper-06, cooper-07, cooper-10, quadri-12, darvish-16, jian-17, jian-18, kawinwanichakij-17}). However, efforts to detect environmental quenching at higher redshift have been hampered for two reasons: first, the practical difficulty  of identifying and spectroscopically confirming protoclusters in the early Universe and second, the observational cost of acquiring the deep multi-passband photometric observations required to make a measurement of the quenched fraction.

One technique which has been successfully employed to identify protocluster systems in the early Universe is to search for ``overdensities'' of, for example, \Halpha\ emitters (HAEs),  Ly$\alpha$ emitters (LAEs) or Lyman-break galaxies \citep[LBGs; \eg][]{steidel-98, lemaux-09, lemaux-14a, dey-16, ouchi-18, toshikawa-18, harikane-19, shi-19, guaita-20, koyama-20}. Another approach has been to target ``signposts''~---~e.g., high-redshift radio galaxies (HzRGs; \citealt{pentericci-97, miley-08, hatch-11a, galametz-12}), quasi-stellar objects (QSOs; \citealt{adams-15}), or dusty star-forming galaxies (DSFGs; \citealt{ivison-01, long-20}). A third technique has been to search for overdensities of DSFGs detected in the far-IR or sub-millimeter \eg\ in  \textit{Herschel Space Telescope}, \textit{Planck Space Telescope}, or South Pole Telescope
surveys  (\citealt{clements-14, planck-XXVII-15, planck-XXXIX-16, greenslade-18, miller-18, cheng-19}).
Examples of protoclusters which have been discovered to date are notable in that they appear to be filled with star-forming galaxies (\citealt{chapman-09, dannerbauer-14, casey-15, hung-16, forrest-17}). Indeed, some authors have suggested that starburst galaxies  may be ubiquitous in protocluster systems \citep{casey-16}.

Here, we report on the discovery of two new protocluster systems, \protoclusterA\ and \protoclusterB, confirmed to be at similar redshift and at a separation of $\sim$35 comoving Mpc. The two systems were discovered not from a dedicated protocluster search but rather during a spectroscopic survey of a sample of ultra-massive galaxies (UMGs; stellar mass $\Mstar > 10^{11.0}~\Mo$) and their environments at $3 <z< 4$. Characterization of the properties of the 16 UMGs spectroscopically confirmed to date from this survey, the ``Massive Ancient Galaxies At z > 3 NEar-infrared'' (\surveyname) survey, has previously been presented in \citealt{forrest-20b} (see also \citealt{forrest-20a}).

The paper is organized as follows: In \S\ref{sec:obs}, we present the target selection, spectroscopic observations, data reduction, and determination of spectroscopic redshifts. In \S\ref{sec:protocluster}, we determine spectroscopic and photometric members of the protocluster systems. In \S\ref{sec:passive}, we calculate rest-frame $UVJ$ colors and quiescent fractions. We discuss our results in \S\ref{sec:disc} and present our main conclusions in \S\ref{sec:conclusions}. We assume $\Omega_{m}=0.3$, $\Omega_\lambda=0.7$, $H_{0}=70$~km~s$^{-1}$~Mpc$^{-1}$, and a Chabrier initial mass function \citep[IMF;][]{chabrier-03} throughout. All magnitudes are on the AB system \citep{oke-83}.

\section{Target Selection and MOSFIRE Observations}
\label{sec:obs}

\subsection{The COSMOS UltraVISTA Field}
\label{ssec:COSMOS}

The COSMOS/UltraVISTA field contains the deepest, highest-quality multi-passband optical, infrared and {\it Spitzer} IRAC imaging available over degree scales. This includes multi-passband imaging taken as part of the COSMOS survey (\citealt{capak-07}),  CFHT-Deep Legacy Survey (\citealt{hildebrandt-09a}),  Subaru Strategic Program (SSP; \citealt{aihara-18}), and UltraVISTA (\citealt{mcCracken-12}) ``ultra-deep stripes,'' providing an exquisite set of photometric measurements in multiple bands, which can be used to estimate photometric redshifts, stellar masses, and rest-frame $UVJ$ colors through spectral energy distribution (SED) modeling. The field is also covered by \textit{GALEX}, \textit{Chandra}, \textit{XMM-Newton}, \textit{Herschel}, SCUBA, and VLA, as well as spectroscopic surveys such as zCOSMOS (\citealt{lilly-07})
and LEGA-C (\citealt{vanderWel-16}).

The unique quality and diversity  of observations in the COSMOS UltraVISTA field has facilitated the discovery of protoclusters using a variety of techniques. These include X-ray emission (\citealt{finoguenov-07, wang-16}), and overdensities in photometric redshift (\eg\ \citealt{Chiang2014, cucciati-18}), distant red galaxies, LAEs, HAEs (\citealt{geach-12}), radio sources (\citealt{daddi-17}), or 3D Ly$\alpha$ forest tomography (\citealt{lee-14a}). Notable spectroscopically-confirmed protoclusters at $z>2$ which have been discovered in the COSMOS UltraVISTA field include systems at $z=2.095$ (\citealt{spitler-12, yuan-14, casey-16, hung-16, tran-17, zavala-19}), $z=2.16$ (\citealt{koyama-20}), $z=2.232$ (\citealt{darvish-20}), $z=2.44$ (``Colossus''; \citealt{lee-16a}, see also \citealt{chiang-15}), $z=2.446$  (``Hyperion''; \citealt{Chiang2014, cucciati-18, diener-13, diener-15}), $z=2.47$ (\citealt{casey-15, casey-16, zavala-19}), $z=2.506$ (CLJ1001; \citealt{wang-16, daddi-17}), $z=2.895$ (\citealt{cucciati-14, cucciati-18}),  $z\sim4.57$ (PCl J1001+0220; \citealt{lemaux-18}), $z\sim5.3$ (\citealt{capak-11a}), and $z=5.667$ (\citealt{pavesi-18}).

\begin{figure*}[htp]
\centering
\includegraphics{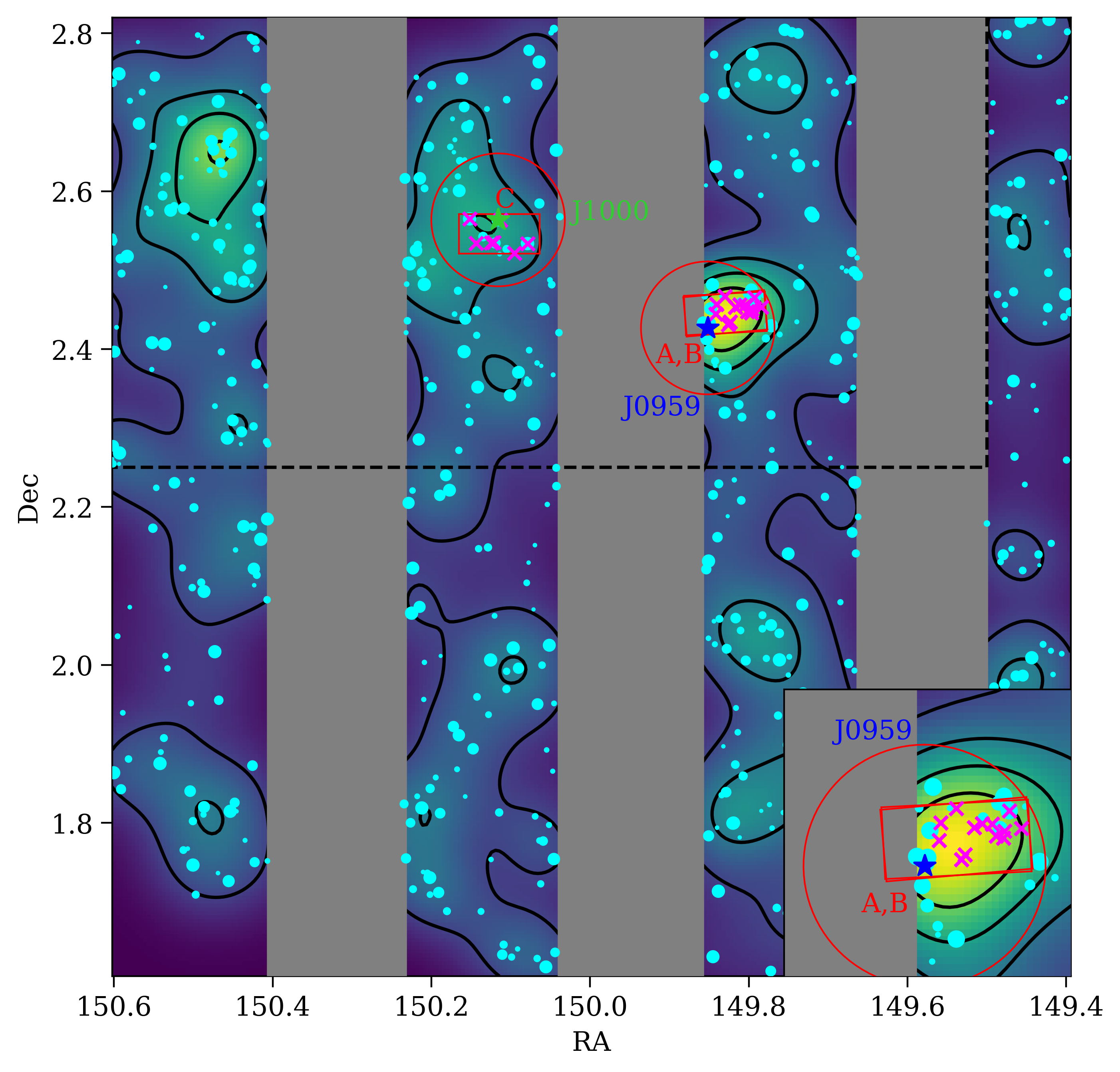}
\caption{
Density distribution of the 550 galaxies in the COSMOS UltraVISTA DR3 catalog (cyan circles) after photometric redshift, stellar mass, $K_{s}$-band magnitude, and probability cuts described in \S\ref{ssec:photz_membership} have been applied. The size of each cyan circle is scaled by the integrated probability, $P$, of the galaxy lying in the redshift interval $z_{\mathrm{J0959}}\pm0.2$ (no galaxies appear in the gray vertical regions because of the ``ultra-deep stripes'' nature of the DR3 survey). 
The solid black contour lines indicate 80\%, 60\%, 40\%, and 20\% of the maximum density (yellow) and were calculated using a Gaussian kernel density estimator as described in \S\ref{ssec:photz_membership}.
The two UMGs, \lUMGA\ (in protocluster \protoclusterA) and \lUMGB\ (in protocluster \protoclusterB), are shown by blue and green stars, respectively (see also \citealt{forrest-20b}). They lie at similar redshift and are separated by $\sim18\amin$ on the sky. Spectroscopically-confirmed protocluster members are shown by magenta crosses and the positions of the three MOSFIRE masks are indicated by red rectangles. The inset at lower right is a zoom-in on the position of  \UMGA, showing the location of masks A and B (which had very similar centers and position angles and are, therefore, somewhat difficult to differentiate). Each of the two red circles has a radius of 10~comoving~Mpc which is approximately equal to the Lagrangian radius. We define the members of protocluster \pA\ to be the 38 galaxies (14 magenta spectroscopic and 24 cyan photometric) which lie within the red circle centered on \UMGA, and the members of protocluster \pB\ to be the 20 galaxies (8 magenta spectroscopic and 12 cyan photometric) which lie within the red circle centered on \UMGB. 
``Field'' galaxies are defined to be the 286 cyan galaxies which lie within the lower half of the DR3 footprint and the upper part of the far right strip, i.e., below and to the right of the black dashed lines (\S\ref{ssec:QF}).
}
\label{fig:densitymap}
\end{figure*}

The work presented here (target selection, SED fits, and photometric analysis) utilizes the COSMOS UltraVISTA Data Release Three (DR3) catalog (Muzzin et al., in prep.; see also \citealt{Marsan2021}), which was constructed using the techniques described in \cite{Muzzin:2013b} for the Data Release 1 (DR1) catalog. The DR3 catalog contains 50 photometric passbands ranging from $u$-band to MIPS 24~$\mu$m, and reaches a total $K_{s}$-band 90\% completeness at $K_{s,{\rm tot}}=24.5$~AB in the 0.84 deg$^{2}$ ``ultra-deep stripes'' area (see Figure~\ref{fig:densitymap}). The DR3 catalog is $\sim1.5$ magnitudes deeper in the $YJHK$ bands than DR1 and also contains the new IRAC SMUVS data (\citealt{Ashby:2018}), which is $\sim1.2$ magnitudes deeper than the S-COSMOS DR1 data (\citealt{sanders-07}).  

Best-fit photometric redshifts and rest-frame colors for galaxies in the catalog were obtained using EAZY \citep{brammer-08} and quantities such as star formation rate, stellar mass, and ages were calculated using FAST \citep{kriek-09}. The photometry was fit to a set of models with exponentially-declining star formation histories ($\mathrm{SFR}\propto~e^{-t/\tau}$), with a time since the onset of star formation ($t$) and a timescale for the decline in the SFR ($\tau$). We used the models of \citet{Bruzual2003} with solar metallicity, a \citet{Calzetti2000} dust law, and assumed a \citet{chabrier-03} IMF. The variables were fit on a grid with $\mathrm{log}(\tau/\rm{yr})$ allowed to range between 7.0 and 10.0, $\mathrm{log}(t/\rm{yr})$ between 7.0 and 10.1, and $\rm{A}_v$ between 0 and 5. The age of a galaxy was limited by the age of the Universe at its redshift.

\subsection{\surveyname\ survey and UMGs \A\ and \B}
\label{ssec:selection}

\lUMGA~(hereafter 179370) and \lUMGB~(hereafter 160748) are two members of a sample of UMGs and their environments at $3 < z < 4$ which we have been targeting for spectroscopic observations using the MOSFIRE spectrograph (\citealt{mclean-10, mclean-12}) on the W. M. Keck Observatory (PI Wilson). This sample of UMGs were selected photometrically from multi-passband optical-infrared catalogs of the COSMOS UltraVISTA (Muzzin et al., in prep.) and VIDEO fields (Annunziatella et al., in prep.). MOSFIRE spectra and stellar population properties (stellar mass, star-formation rate, star-formation history, quiescence) of the 16 \surveyname\ UMGs which have been spectroscopically confirmed to date were presented in \citet{forrest-20b}, including the discovery of the most massive spectroscopically-confirmed quiescent UMG yet confirmed at $z>3$ (\citealt{forrest-20a}). A key goal of the  \surveyname\ survey is to utilize MOSFIRE's powerful multiplexing capabilities in combination with the uniquely deep and extensive DR3 (and VIDEO) catalogs to characterize not only each UMG but also its environment.

\UMGA\ and \B\ were first identified as candidate UMGs at $3 < z< 4$ by \cite{Marchesini:2010}, based on analysis of the 
NEWFIRM Medium-Band Survey (NMBS; \citealt{whitaker-11}). \citet{marsan-17} presented spectroscopic confirmation of \UMGA\  (ID ``C1-15182'') by means of the \OIII\ doublet. Based on X-ray observations and line flux ratios, \UMGA\ contains a powerful AGN. \UMGA\ is estimated to have a star formation rate (SFR) of less than $100~\Mo~{\rm yr}^{-1}$  based on UV-to-FIR SED-fitting (\citealt{marsan-17}) or alternatively, less than $15~\Mo~{\rm yr}^{-1}$ based on its \Hbeta\ line flux (\citealt{forrest-20b}).

Spectroscopic confirmation of \UMGB, a post-starburst (PSB) galaxy, was first presented in \citet{Marsan2015}, and it has since been studied further extensively [identified by ID ``C1-23152'' in \citet{Marsan2015} and  \cite{saracco-20}, and ``COS-DR3-160748'' in \cite{forrest-20b}]. Detailed analysis of its UV-to-FIR SED revealed that most of its stars formed at $z>4$ in a highly dissipative, intense, and short burst of star formation. Based on its emission line ratios, it also contains a powerful AGN. \UMGB\ has a star formation rate of less than $10~\Mo~{\rm yr}^{-1}$, and has negligible dust extinction (\citealt{marsan-17, forrest-20b, saracco-20}). The bulk of the stars in \UMGB\ appear to have super-solar metallicity and the dynamical mass estimated from the stellar velocity dispersion is consistent with the stellar mass derived from SED fitting (\citealt{saracco-20}).

\begin{figure*}[!htb]
\centering
\includegraphics{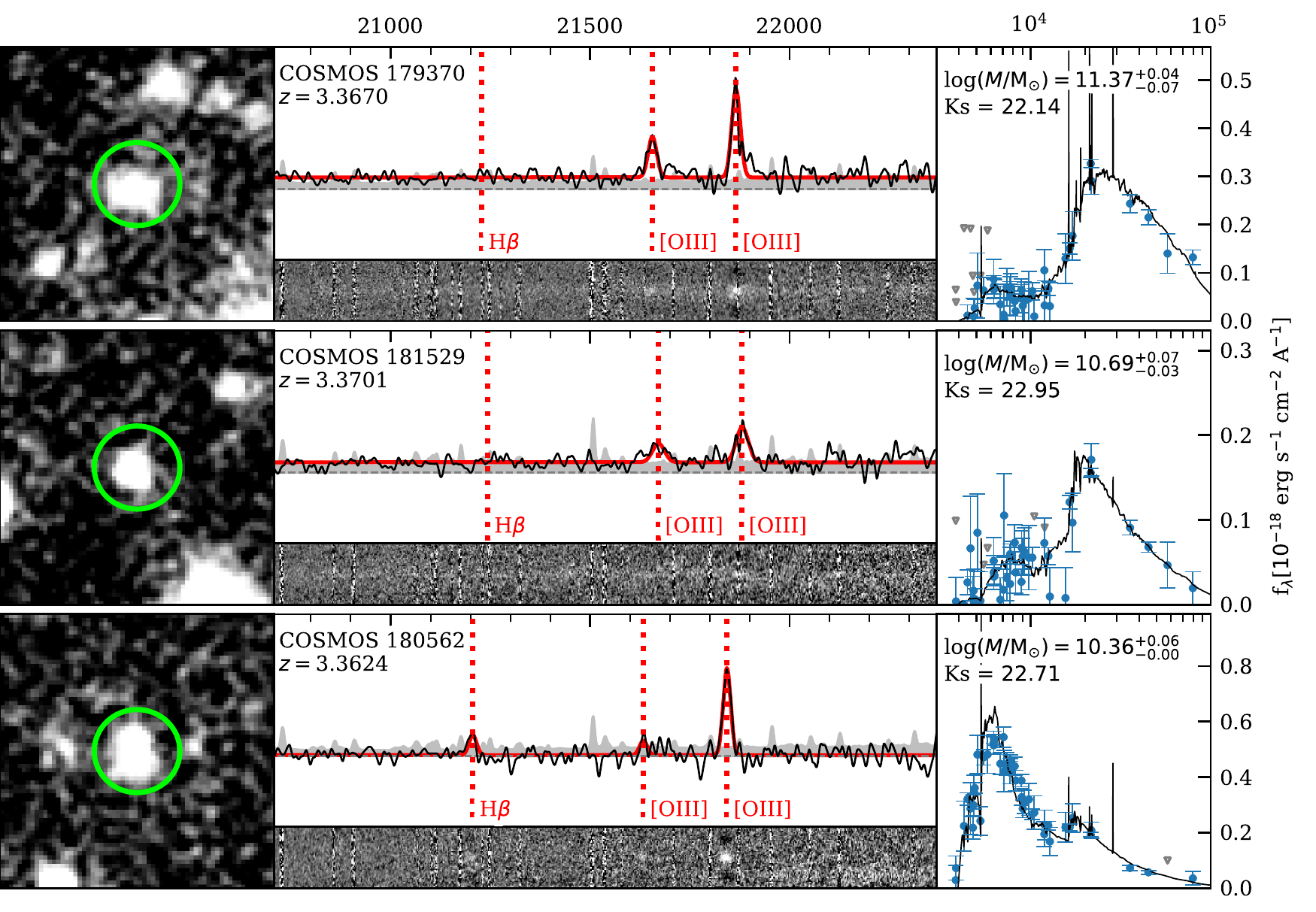}
\caption{$K_{s}$-band image (left),  MOSFIRE  1D $K$-band spectra (upper center), MOSFIRE 2D $K$-band spectra (lower center) and the SED (right) of spectroscopically-confirmed members (the first three members of Table~\ref{table:allspeczs} are shown here; all 22 members are shown in Figure~\ref{fig:spectra}). The black solid line shows the spectrum smoothed over 5 pixels weighted by the inverse variance. The light gray shading shows the magnitude of the error spectrum and the horizontal dashed dark grey line indicates where the flux is zero. The solid red line is the best-fit six-parameter model described in \S\ref{ssec:redshift_determination}. The vertical red dotted lines show the position at which  \Hbeta\ and \OIII\ doublet emission lines would appear at the spectroscopic redshift of each galaxy. The photometric fluxes and their  $1\sigma$ errors are shown on the right in blue, with the best-fit SED shown in black. For those bands for which the flux is negative, the 3$\sigma$ upper limit is shown as a gray downward-pointing triangle.
}
\label{fig:spectra_short}
\end{figure*}

\subsection{MOSFIRE Spectroscopic Observations and Data Reduction}

As shown in  Figure~\ref{fig:densitymap}, two $K$-band masks (A and B) centered on \UMGA\ were observed in November 2017 and one $K$-band mask (C) centered on \UMGB\ was observed in March 2019. Filler slits were placed on DR3 targets with photometric redshift $z_{\rm phot}\pm0.3$ of the UMG, with priority given to galaxies with total $K_{s}$-band magnitude brighter than $K_{s, {\rm tot}}=23.0$.  Exposure times ranged between 3500s and 9300s. The observations are summarized in Table~\ref{table:overview}. 

\begin{table}[htbp]
\centering
\caption{\label{table:overview} Overview of Observations}
\begin{tabular}{lccc}
\hline
\hline
Mask         &Observation date                  & Exposure time (s)   & Seeing (FWHM)        \\
\hline
A              & 2017 Nov 21		&9304.8 & 0.75''\\
B            & 2017 Nov 22		&7873.3 & 0.73''\\
\hline
C		&2019 Mar 18	&3578.0 & 0.61''\\
\hline
\end{tabular}
\end{table}

We began reduction by running the MOSFIRE Data Reduction Pipeline\footnote{https://github.com/Keck-DataReductionPipelines/MosfireDRP} (DRP) to obtain 2D target and error spectra. The DRP constructs a pixel flat image, identifies slits, removes thermal contamination, performs wavelength calibration using sky lines and neon arc lamps, removes sky background, and rectifies the spectrum.

We extracted the 1D spectra with a modified version of the MOSFIRE DRP designed to perform telluric corrections and mask skylines \citep{forrest-20b}. By visually inspecting the 2D spectrum, we determined whether stellar continuum or an emission feature was present. When stellar continuum was present, we collapsed the 2D spectrum along the wavelength axis to identify the location of the trace. A Gaussian was then fit to the collapsed spectrum and used as the weighting for optimal extraction \citep{horne-86}. When only an emission feature was present, we collapsed the spectrum along the limited portion of wavelength space containing the emission feature and between the nearest sky lines. We then applied a telluric correction using spectra of the science calibration stars, and masked regions with contamination from bright skylines to obtain a final 1D extracted spectrum \citep[see][]{forrest-20b}. For objects which appeared in both masks A and B, we weighted the extracted 1D spectra and the noise spectra by the inverse variance and co-added them.

\subsection{Redshift Determination}
\label{ssec:redshift_determination}

For an emission line galaxy at $z\sim3.37$, \Hbeta\ and the \OIII\ doublet fall in the observed $K$ band. In order to obtain a spectroscopic redshift we fit a three-Gaussian model to the emission features in the 1D extracted spectrum with six free parameters: redshift, the fluxes of the three lines, a line width (identical for all three lines), and a constant flux offset to account for the stellar continuum. When fitting the model to the \OIII\ doublet, we fit the fluxes of the emission lines independently, but the best fit generally resulted in a line ratio very close to the expected ratio of 1:3. We obtained spectroscopic redshifts for 14 galaxies including  \UMGA\ from masks A and B, and 8 galaxies including \UMGB\ from mask C (Table \ref{table:allspeczs}).

\begin{figure*}[htp]
\centering
\caption{As for Figure~\ref{fig:spectra_short} but for all 22 members, shown in the same order as in Table~\ref{table:allspeczs}.} 
\label{fig:spectra}
\begin{tabular}{c}
\\
\includegraphics[width=\textwidth]{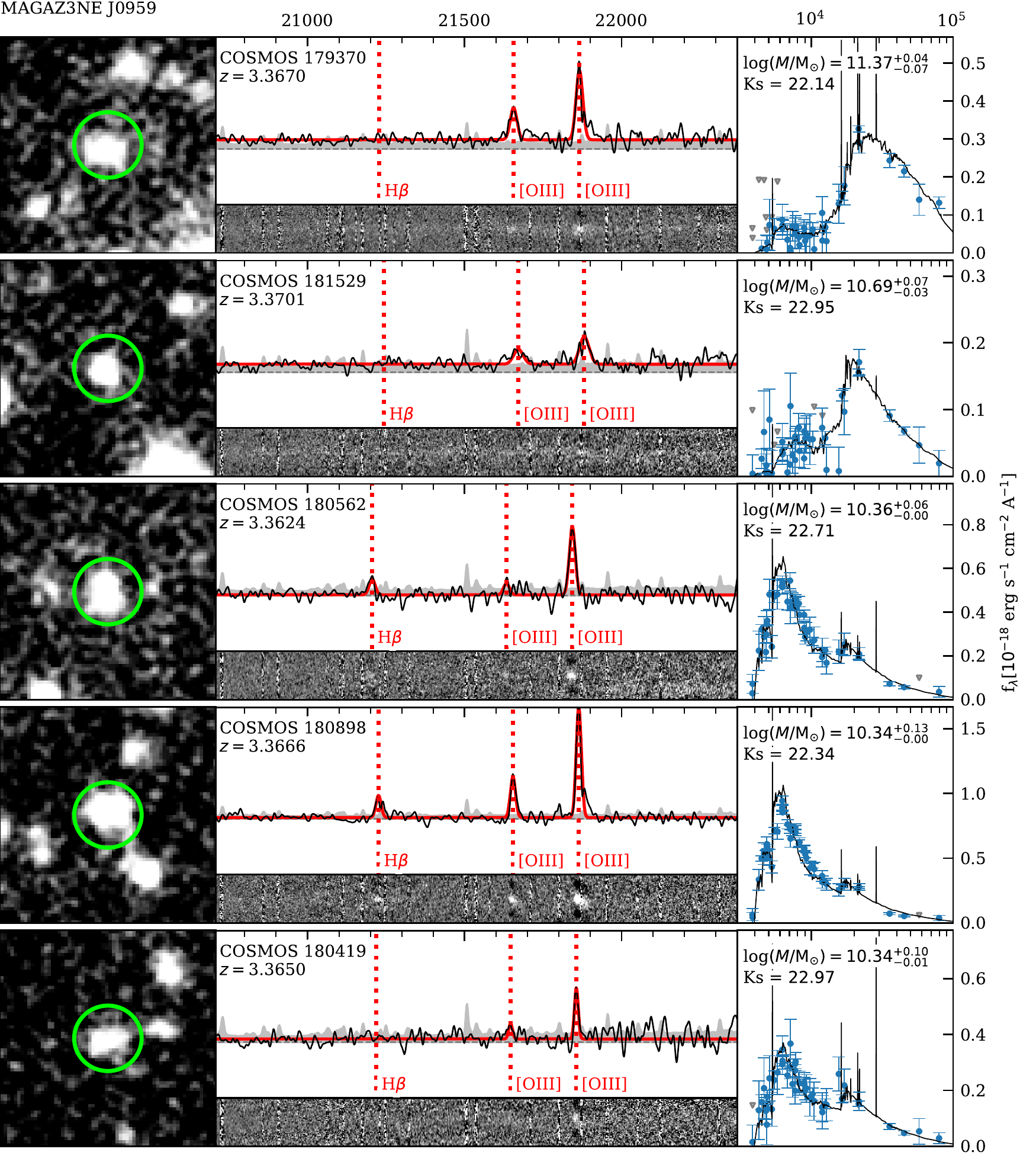}

\end{tabular}
\end{figure*}

Figure~\ref{fig:spectra_short} shows the $K_{s}$-band images (left), 2D and 1D $K$-band spectra (center), and SEDs (right) 
for the first three galaxies in Table~\ref{table:allspeczs} (Figure~\ref{fig:spectra} shows same but for all galaxies in Table~\ref{table:allspeczs}). The black solid line shows the 1D spectrum smoothed over five pixels ($\sim 11$ \AA), weighted by the inverse variance. The light gray shading shows the error spectrum. The solid red line is the best-fit six-parameter model described above, while the dotted red vertical lines show the wavelengths corresponding to \Hbeta\ and \OIII\ at the best-fit spectroscopic redshift, $z_{\rm spec}$.

The uncertainty on each spectroscopic redshift was obtained by adding statistical and systematic uncertainties in quadrature.
In order to calculate the statistical uncertainty for each spectrum, 1000 simulated spectra were created by perturbing the flux at each wavelength of the observed spectrum by a Gaussian random amount with the standard deviation set by the level of the $1\sigma$ error spectrum. The 1000 simulated spectra were then fit to obtain a distribution of values for the redshift. Upper and lower $1\sigma$  confidence limits (\ie\ the statistical uncertainty) were obtained by integrating the redshift probability distribution to find the 16th and 84th percentile values. On average, the statistical uncertainty obtained was $\delta z \sim 0.0002$.

The systematic error on the redshift was calculated by multiplying the spectral dispersion (2.17\AA/pixel) by the pixel resolution (2.78 pixels), to obtain the spectral resolution ($6.03$\AA). At $z=3.37$ this spectral resolution corresponds to $\delta z \sim 0.0012$. In every case, the systematic uncertainty dwarfed the statistical uncertainty.

The left panel of Figure~\ref{fig:speczhist_both} shows the excellent agreement between the spectroscopic and photometric redshifts. The blue and green stars indicate \UMGA\ and \UMGB, and the blue and green squares show galaxies with spectroscopic redshifts near \UMGA\ (masks A and B) and \UMGB\ (mask C), respectively. Members with broader photometric redshift probability distributions have larger photometric redshift uncertainties.  The median of the scatter in $| z_\mathrm{spec} - z_\mathrm{phot}| / (1+z_\mathrm{spec})$ is 0.0058.

We obtained spectroscopic redshifts for 22 galaxies in the redshift range $3.2 < z< 3.5$ (including both UMGs). In order to derive more accurate estimates of stellar mass, star formation rate, and age for those 22 galaxies shown in Table \ref{table:allspeczs}, we fixed $z=z_{\rm spec}$ and then reran FAST (\citealt{kriek-09}) on the DR3 catalog using the same parameters as in \S \ref{ssec:COSMOS}.

\section{Protocluster Membership}
\label{sec:protocluster}

\subsection{Spectroscopic Members}
\label{ssec:specz_membership}

\begin{figure*}[htbp]
\includegraphics[width=\linewidth]{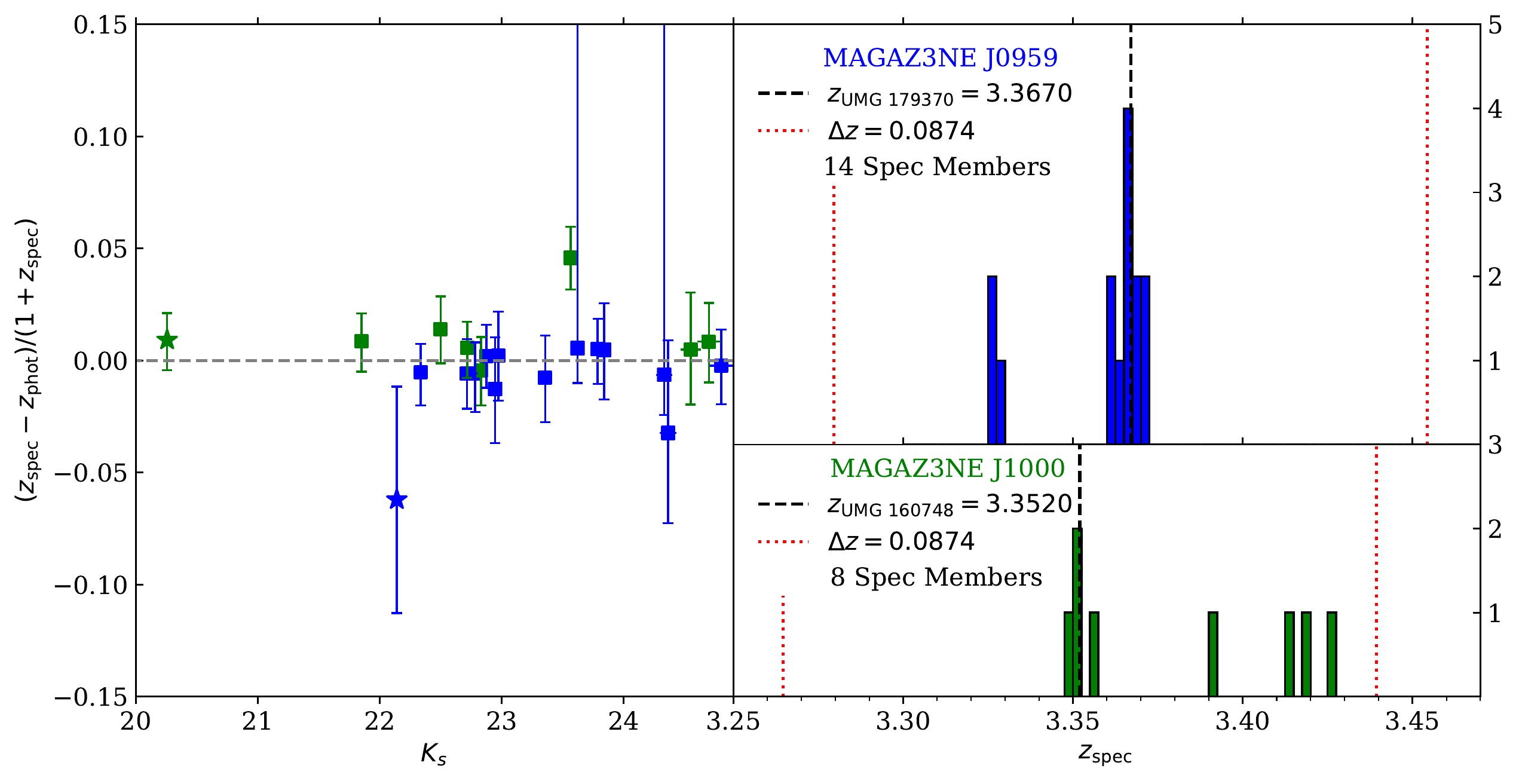}
\caption{{\bf Left:} 
Blue and green symbols show the 22 spectroscopic members of protocluster \pA\ (upper right) and protocluster \pB\ (lower right). The blue and green stars indicate \UMGA\ and \UMGB, respectively. There is excellent agreement between the spectroscopic and photometric redshifts for the 22 spectroscopic members (members with broader photometric redshift probability distributions have larger photometric redshift uncertainties). {\bf Right:} Histograms of the spectroscopic members shown at left (with properties summarized in Table~\ref{table:allspeczs}). Upper panel shows galaxies with spectroscopic redshifts close to \UMGA\  (masks A and B in Table~\ref{table:overview}) and lower panel shows \UMGB\  (mask C in Table~\ref{table:overview}).  The redshifts of the two UMGs are shown by the dashed black lines. We consider the 22 galaxies with velocities within $\pm6000$~km~s$^{-1}$ ($ \Delta z=0.0874$) of each of the UMGs~(red dashed lines) to be spectroscopic protocluster members. Those galaxies are indicated by magenta crosses in Figure~\ref{fig:densitymap}. 
}
\label{fig:speczhist_both}
\end{figure*}

The right panel of Figure~\ref{fig:speczhist_both} shows a histogram of the galaxies with spectroscopic redshifts close to that of \UMGA\ (upper; blue) and to that of \UMGB\ (lower; green). The redshifts of the UMGs are shown by the dashed black lines.  We consider the 22 galaxies with line-of-sight velocities within $\pm6000$~km~s$^{-1}$ ($ \Delta z = 0.0874$) of each of the UMGs (red dashed lines) to be spectroscopic members. There are 14 spectroscopic members (including \UMGA) of protocluster \protoclusterA\  (hereafter J0959) and 8 spectroscopic members (including \UMGB) of protocluster \protoclusterB\ (hereafter J1000). In naming each protocluster, the R.A. and decl. was chosen to coincide with the coordinates of its UMG. The positions of the 22 spectroscopic protocluster members are shown by  magenta crosses in Figure~\ref{fig:densitymap}. The $\pm6000$~km~s$^{-1}$ velocity cut used here is similar to that used to determine spectroscopic membership for other high-redshift protoclusters (\citealt{lemaux-14a}), and is well matched to the redshift extent of simulated protostructures  (\citealt{chiang-13, muldrew-15}).

Using the biweight location estimator (\citealt{beers-90}), we determined the mean redshift of the 14 spectroscopically confirmed members of \pA\ to be $z = 3.3665^{+0.0009}_{-0.0012}$ and the mean redshift of the 8 members of \pB\ to be $z = 3.3801^{+0.0213}_{-0.0281}$. The uncertainties on the mean redshifts were calculated using bootstrapping.

\begin{deluxetable*}{cccccccccc}
\tabletypesize{\footnotesize}
\tablecolumns{10} 
\tablewidth{\textwidth}
 \tablecaption{Properties of Spectroscopic Members of \pA\ (upper) and \pB\ (lower), ordered by stellar mass
 \label{table:allspeczs}}
 \tablehead{
 \colhead{ID} \vspace{-0.2cm}& \colhead{Mask} & \colhead{R.A.} &  \colhead{Decl.}& \colhead{$K_s$} &   \colhead{$z_{\rm spec}$} & \colhead{C$^{a}$}& \colhead{Stellar Mass}  & \colhead{Age} & \colhead{SFR}\\
 \colhead{} & \colhead{}& \colhead{}& \colhead{}& \colhead{}& \colhead{}& \colhead{}&  \colhead{\logM} & \colhead{log(yr)} & \colhead{log(\Mo yr$^{-1}$)}}
 \startdata 
\vspace{0.1cm} 179370$^{b}$ & A, B & $ 9^{\mathrm{h}}59^{\mathrm{m}}24.3936^{\mathrm{s}}$  &  $+2^\circ25'36.5117''$ & 22.14 & $3.3670\pm0.0012$ & 1 & $11.37_{-0.07}^{+0.04}$ & $9.0_{-0.1}^{+0.1}$ & $0.50_{-0.60}^{+0.59}$\\
\vspace{0.1cm}181529 & A, B & $ 9^{\mathrm{h}}59^{\mathrm{m}}10.2576^{\mathrm{s}}$  &  $+2^\circ27'54.0562''$ & 22.95 & $3.3701\pm0.0013$ & 1 & $10.69_{-0.04}^{+0.06}$ & $8.7_{-0.1}^{+0.1}$ & $-0.35_{-0.65}^{+0.81}$\\
\vspace{0.1cm}180562 & A & $ 9^{\mathrm{h}}59^{\mathrm{m}}11.2104^{\mathrm{s}}$  &  $+2^\circ26'45.7015''$ & 22.71 & $3.3624\pm0.0012$ & 1 & $10.36_{-0.00}^{+0.05}$ & $8.4_{-0.1}^{+0.1}$ & $1.00_{-0.08}^{+0.01}$\\
\vspace{0.1cm}180898 & A & $ 9^{\mathrm{h}}59^{\mathrm{m}}8.2152^{\mathrm{s}}$  &  $+2^\circ27'10.3471''$ & 22.34 & $3.3666\pm0.0012$ & 1 & $10.34_{-0.02}^{+0.11}$ & $8.3_{-0.0}^{+0.1}$ & $1.08_{-0.04}^{+0.02}$\\
\vspace{0.1cm}180419 & A & $ 9^{\mathrm{h}}59^{\mathrm{m}}21.9552^{\mathrm{s}}$ & $+2^\circ26'40.1554''$ & 22.97 & $3.3650\pm0.0012$ & 2 & $10.34_{-0.01}^{+0.09}$ & $8.5_{-0.1}^{+0.0}$ & $0.54_{-0.13}^{+0.48}$\\
\vspace{0.1cm}180910 & A & $ 9^{\mathrm{h}}59^{\mathrm{m}}16.1016^{\mathrm{s}}$  &  $+2^\circ27'12.1205''$ & 22.78 & $3.3676\pm0.0012$ & 1 & $10.32_{-0.06}^{+0.03}$ & $8.5_{-0.1}^{+0.0}$ & $1.33_{-0.24}^{+0.19}$\\
\vspace{0.1cm}179810 & A & $ 9^{\mathrm{h}}59^{\mathrm{m}}17.6304^{\mathrm{s}}$  &  $+2^\circ26'4.3588''$ & 23.36 & $3.3270\pm0.0012$ & 1 & $10.25_{-0.03}^{+0.07}$ & $8.3_{-0.0}^{+0.2}$ & $1.24_{-0.12}^{+0.35}$\\
\vspace{0.1cm}181634 & A & $ 9^{\mathrm{h}}59^{\mathrm{m}}19.1352^{\mathrm{s}}$  &  $+2^\circ27'59.7582''$ & 22.87 & $3.3696\pm0.0012$ & 1 & $10.21_{-0.03}^{+0.06}$ &  $8.5_{-0.1}^{+0.0}$ & $1.02_{-0.04}^{+0.03}$\\
\vspace{0.1cm}181058 & B & $ 9^{\mathrm{h}}59^{\mathrm{m}}21.7272^{\mathrm{s}}$  &  $+2^\circ27'23.5634''$ & 23.84 & $3.3287\pm0.0013$ & 3 & $10.13_{-0.11}^{+0.03}$ & $8.4_{-0.1}^{+0.1}$ & $1.04_{-0.48}^{+0.20}$\\
\vspace{0.1cm}181039 & A & $ 9^{\mathrm{h}}59^{\mathrm{m}}14.7624^{\mathrm{s}}$  &  $+2^\circ27'22.3693''$ & 23.62 & $3.3661\pm0.0012$ & 2 & $10.05_{-0.05}^{+0.04}$ & $8.4_{-0.1}^{+0.0}$ & $0.69_{-0.21}^{+0.34}$\\
\vspace{0.1cm}180737 & B & $ 9^{\mathrm{h}}59^{\mathrm{m}}11.0640^{\mathrm{s}}$  &  $+2^\circ27'2.5391''$ & 23.79 & $3.3622\pm0.0012$ & 1 & $9.86_{-0.08}^{+0.11}$ & $8.6_{-0.1}^{+0.2}$ & $1.03_{-0.13}^{+0.10}$\\
\vspace{0.1cm}180577 & A, B & $ 9^{\mathrm{h}}59^{\mathrm{m}}12.4320^{\mathrm{s}}$  &  $+2^\circ26'50.5907''$ & 24.33 & $3.3707\pm0.0013$ & 3 & $9.86_{-0.14}^{+0.07}$ & $8.6_{-0.2}^{+0.1}$ & $0.58_{-0.43}^{+0.30}$\\
\vspace{0.1cm}179570 & B & $ 9^{\mathrm{h}}59^{\mathrm{m}}18.2304^{\mathrm{s}}$  &  $+2^\circ25'51.8840''$ & 24.36 & $3.3272\pm0.0012$ & 3 & $9.72^{+0.15}_{-0.05}$ & $8.5^{+0.1}_{-0.1}$ & $-0.08^{+0.67}_{-0.21}$\\
\vspace{0.25cm}180993 & A & $ 9^{\mathrm{h}}59^{\mathrm{m}}13.0872^{\mathrm{s}}$  &  $+2^\circ27'20.5978''$ & 24.80  & $3.3646\pm0.0012$ & 1 & $8.99_{-0.14}^{+0.17}$ & $8.1_{-0.4}^{+0.2}$ & $0.36_{-0.28}^{+0.1}$\\
\hline
\hline
\vspace{0.1cm}160748$^{b}$ & C & $10^{\mathrm{h}} 0^{\mathrm{m}}27.8112^{\mathrm{s}}$  &  $+2^\circ33'49.2289''$ & 20.26 & $3.3520\pm0.0012$ & 1 & $11.47_{-0.03}^{+0.03}$ & $8.6_{-0.1}^{+0.0}$ & $0.49_{-0.03}^{+0.03}$\\
\vspace{0.1cm}158792 & C & $10^{\mathrm{h}}  0^{\mathrm{m}}18.8880^{\mathrm{s}}$  &  $+2^\circ32'1.0410''$ & 22.50 & $3.4136\pm0.0012$ & 1 & $10.68_{-0.05}^{+0.03}$ & $8.6_{-0.1}^{+0.0}$ & $1.39_{-0.06}^{+0.25}$\\
\vspace{0.1cm} 160752 & C & $10^{\mathrm{h}}  0^{\mathrm{m}}36.4296^{\mathrm{s}}$  &  $+2^\circ33'53.6620''$ & 21.85 & $3.4261\pm0.0012$ & 1 & $10.53_{-0.05}^{+0.05}$ & $8.3_{-0.1}^{+0.0}$ & $1.72_{-0.15}^{+0.09}$\\
\vspace{0.1cm}158809 & C & $10^{\mathrm{h}}  0^{\mathrm{m}}34.5120^{\mathrm{s}}$  &  $+2^\circ32'2.2290''$ & 22.83 & $3.3910\pm0.0012$ & 1 & $10.34_{-0.02}^{+0.03}$ & $8.4_{-0.1}^{+0.0}$ & $0.64_{-0.12}^{+0.02}$\\
\vspace{0.1cm}158027 & C & $10^{\mathrm{h}}  0^{\mathrm{m}}22.8432^{\mathrm{s}}$  &  $+2^\circ31'17.6844''$ & 22.72 & $3.4176\pm0.0012$ & 3 & $10.30_{-0.02}^{+0.07}$ & $8.6_{-0.1}^{+0.1}$ & $1.02_{-0.04}^{+0.02}$\\
\vspace{0.1cm}158806 & C & $10^{\mathrm{h}}  0^{\mathrm{m}}29.9016^{\mathrm{s}}$  &  $+2^\circ32'3.7946''$ & 23.56 & $3.3509\pm0.0012$ & 2 & $10.02_{-0.04}^{+0.07}$ & $8.8_{-0.1}^{+0.0}$ & $0.90_{-0.09}^{+0.01}$\\
\vspace{0.1cm}158838 & C & $10^{\mathrm{h}}  0^{\mathrm{m}}29.1840^{\mathrm{s}}$  &  $+2^\circ32'5.3916''$ & 24.55 & $3.3493\pm0.0012$ & 2 & $9.54_{-0.11}^{+0.07}$ & $8.5_{-0.2}^{+0.1}$ & $0.35_{-0.31}^{+0.05}$\\
\vspace{0.25cm}160549 & C & $10^{\mathrm{h}} 0^{\mathrm{m}}27.0312^{\mathrm{s}}$  &  $+2^\circ33'48.3826''$ & 24.70 & $3.3572\pm0.0012$ & 1 & $9.42_{-0.12}^{+0.11}$ & $8.5_{-0.2}^{+0.2}$ & $0.58_{-0.24}^{+0.18}$\\
\hline
\enddata

\tablenotetext{a}{
The spectroscopic redshift confidence level was assigned based on the number of emission lines observed. A spectrum where two emission lines were observed, e.g., \Hbeta, or one or both lines of the \OIII\ doublet, was assigned a confidence level of 1, a spectrum where a single high S/N emission line ($\rm{S/N} \geq 3$) was observed was assigned a confidence level of 2, and a spectrum where a single low S/N ($\rm{S/N} < 3$) emission line was observed was assigned a confidence level of 3.}
\tablenotetext{b}{
UMG (primary target)}
\end{deluxetable*}

\subsection{Photometric Members}
\label{ssec:photz_membership} 

There are 236,196 objects in the DR3 catalog. In order to determine photometric membership for the two protoclusters, we began by fitting each of the galaxies in the DR3 catalog with EAZY (\citealt{brammer-08}) to derive a best-fit photometric redshift, $z_{\rm peak}$, and with FAST (\citealt{kriek-09}) to derive the stellar mass and star formation rate. Simulations have shown that the DR3 catalog is 90\% complete down to a total $K_{s}$-band magnitude of $K_{s,{\rm tot}}=24.5$, and that this corresponds to a 95\% completeness above a  stellar mass of  \logM~$=10.5$ at $z\sim4$ (Muzzin et al., in prep.). We, therefore, selected those galaxies with $K_{s}\leq24.5$ and \logM~$\geq10.5$, and also made an initial photometric selection of $2.75<z_{\rm peak}<4$. There were 1,279 galaxies which satisfied those three criteria.

Next we wished to select the members of the ``slice'' centered on \pA. In determining membership, photometric analyses often simply select galaxies with $z_{\rm phot}$ within a given redshift range. However, this does not take account of the fact that each galaxy has a different redshift probability distribution function, $p(z)$. Therefore here, to account for the diversity in the $p(z)$ distributions, we instead adopted a probabilistic selection applied to each galaxy in turn. We integrated $p(z)$ for each galaxy using the redshift of protocluster \pA~as the fiducial central redshift and the median photometric uncertainty of the sample of 1,279 galaxies $\Delta z_{\rm phot, sample}$, as the lower and upper limits.

\begin{equation}
\label{eq:1}
P = \frac{\int_{z_{\rm J0959}-\Delta z_{\rm phot, sample}}^{z_{\rm J0959}+\Delta z_{\rm phot, sample}}{p(z)dz}}{\int_0^{\infty}p(z)dz}\\
\end{equation}

Galaxies with an integrated probability $P$ in excess of a threshold probability $P_{\rm thresh}$ will be considered members of the protocluster redshift slice. We explain below in detail how we calculated the actual values of $\Delta z_{\rm phot, sample}$ and $P_{\rm thresh}$ ($\Delta z_{\rm phot, sample}=0.2$ and  $P_{\rm thresh}=0.17$). However, we note that we utilized different values of $\Delta z_{\rm phot, sample}$ (in the range $0.10 \leq \Delta z_{\rm phot, sample} \leq 0.30$) and $P_{\rm thresh}$ (in the range $0.15 \leq P_{\rm thresh} \leq 0.5$) and any choice of value in those ranges had minimal impact on the number of protocluster members or the quiescent fraction results presented in \S~\ref{ssec:QF}.

In general, the precision of any sample of photometric redshifts is often expressed as a percentage, $\Delta z_{\rm phot, sample} / (1+z_{\rm phot})$. For this sample of 1,279 galaxies, we calculated the median percentage to be $4.5\%$ by determining $\Delta z_{\rm phot, galaxy}/(1+z_{\rm phot})$ for each galaxy using the the 16th and 84th percentile values of its $p(z)$ distribution output by EAZY as that galaxy's photometric redshift uncertainty $\Delta z_{\rm phot, galaxy}$. In other words, the median photometric redshift uncertainty expressed as a function of redshift is $\Delta z_{\rm phot, sample} = 0.045 (1+z_{\rm phot})$, and at the redshift of \pA, $\Delta z_{\rm phot, sample} \sim 0.2$.
Having calculated the value of $\Delta z_{\rm phot, sample}$ (used in the integration limits in Equation~\ref{eq:1}), we then calculated the integrated probability $P$ for each galaxy using Equation~\ref{eq:2}.

\begin{eqnarray} \label{eq:2}
P &=& \frac{\int_{z_{\rm J0959}-0.2}^{z_{\rm J0959}+0.2}{p(z)dz}}{\int_0^{\infty}p(z)dz}\\
&=& \frac{\int_{3.167}^{3.567}{p(z)dz}}{\int_0^{\infty}p(z)dz} \nonumber
\end{eqnarray}

To determine the threshold value, $P_{\rm thresh}$, we consider a hypothetical galaxy which has a Gaussian $p(z)$ with an uncertainty of three times the median photometric redshift uncertainty, $\Delta z_{\rm phot, sample}$. We would like this galaxy to fall just at the $P_{\rm thresh}$ limit for inclusion in the redshift slice. To achieve this we set the photometric redshift of this hypothetical galaxy such that the redshift of \pA~fell at this galaxy's photometric redshift uncertainty (i.e.,  $z_{\rm phot} \pm \Delta z_{\rm phot, galaxy} = z_{\rm J0959}$). By applying Equation \ref{eq:2} to this hypothetical galaxy, we obtained $P=0.17$, which we then adopt to be the threshold probability $P_{\rm thresh}$. Each of the 1,279 galaxies in our sample with $P \geq P_{\rm thresh}$ are considered to be members of the redshift slice. We note that we experimented with using different values of $P_{\rm thresh}$ (in the range $0.15 \leq P_{\rm thresh} \leq 0.5$) but once again found any choice of value in this range to have minimal impact on the number of protocluster members or the quiescent fraction results presented in \S~\ref{ssec:QF}.

The cyan circles in Figure~\ref{fig:densitymap} show the remaining 550 galaxies which have $P\geq0.17$, each with its size scaled by its $P$ value (galaxies which are spectroscopic members are assigned $P=1$). The smoothed density map in Figure \ref{fig:densitymap} was generated by applying a Gaussian kernel density estimator to the 550 galaxies weighted by their $P$ value, with maximal density colored yellow, and contours drawn at $20\%$ intervals of the maximum value. We chose a kernel bandwidth of 7.7' ($\sim15$ comoving Mpc), which approximately corresponds to the predicted physical size of a massive protocluster at this redshift. The contours and smoothed density map, which are shown in Figure~\ref{fig:densitymap}, have been truncated at the edges of the ultra-deep stripes.

Simulations have shown that 10 comoving Mpc is approximately equal to the Lagrangian radius at $z\sim3.37$ (\citealt{Chiang:2017}). As a final step in determining photometric membership, we retained only those galaxies within a radius of 10 comoving Mpc from each UMG. This selection resulted in a total of 26 photometric members for \pA\ (two photometric members of which, \UMGA\ and galaxy ID ``181529,'',were also independently spectroscopically confirmed). This selection also resulted in a total of 15 photometric members for \pB\ (three photometric members of which, \UMGB\ and galaxy IDs ``158792'' and ``160752,'' were also independently spectroscopically confirmed). We note that most of the spectroscopically-confirmed members of the two protoclusters are emission-line galaxies with stellar masses below the completeness limit of the photometric catalog (Table \ref{table:overview}). As a result, these galaxies were not identified as photometric members despite otherwise having photometric redshifts consistent with membership.

\begin{figure*}[htb]
\centering
\includegraphics[width=\textwidth]{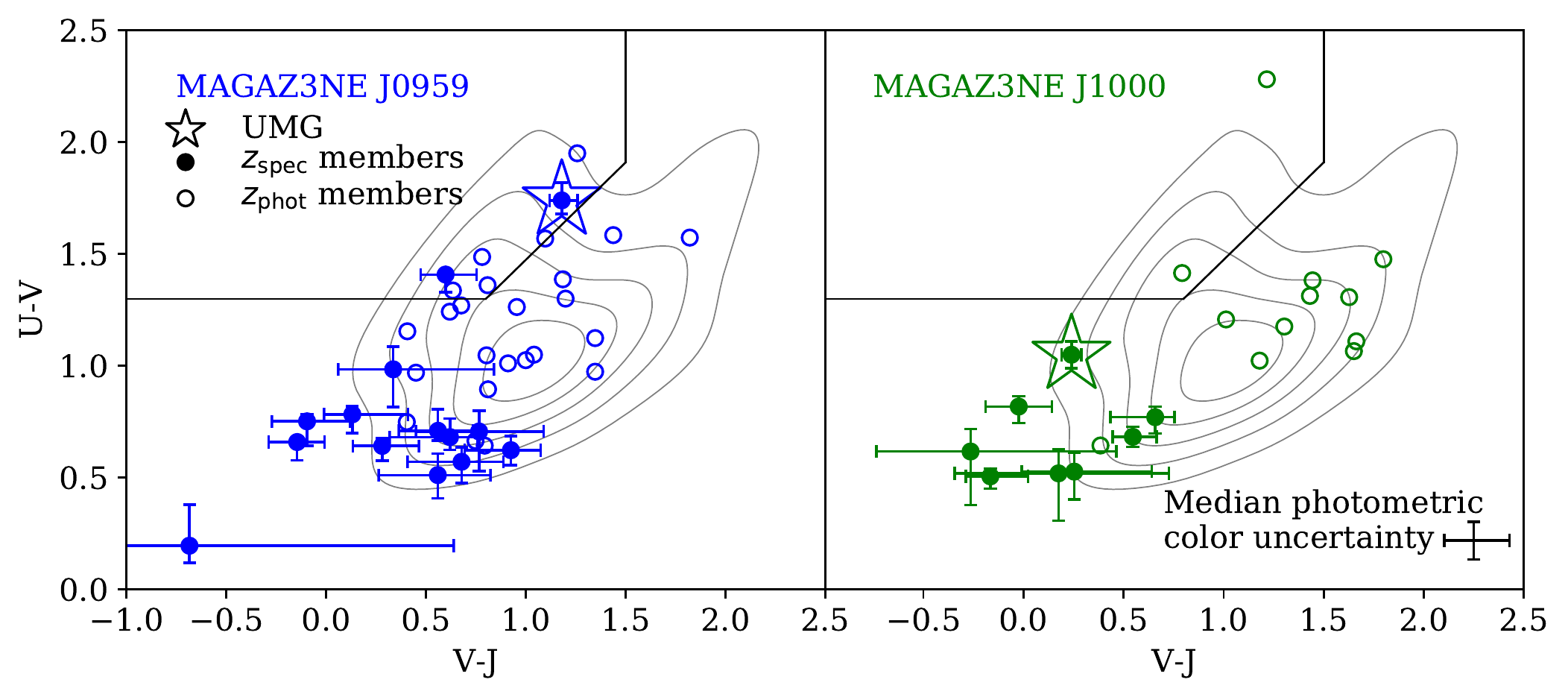}
\caption{
$UVJ$ color-color diagram for \pA\ (left) and \pB\ (right). \UMGA\ is shown by the blue star and \UMGB\ by the green star. The 13 (7) additional spectroscopic members and 24 (12) photometric members of  \pA\ (\pB) are shown by solid and open blue (green) circles. The black cross at bottom right shows the median uncertainty in the colors of the photometric members. 
Note that most of the spectroscopically-confirmed members of the two protoclusters were not identified as photometric members because they are emission-line galaxies which fall below the stellar mass completeness limit (\logM~$=10.5$) applied to the DR3 catalog. Only one spectroscopically-confirmed member, other than the UMGs, has a stellar mass greater than
\logM~$ = 10.5$ (ID ``81529'' in Table~\ref{table:allspeczs}). The contours show the field sample (defined in \S\ref{ssec:QF}), and the wedge defined by the solid black lines shows the quiescent galaxy  selection criteria proposed by \cite{whitaker-11}.
Interestingly, as shown in the left panel, \UMGA\ (and galaxy 181529) are $UVJ$ quiescent. In contrast, because \UMGB's star formation quenched rather recently and abruptly 
(within the last few hundred Myr; \citealt{Marsan2015, marsan-17, forrest-20b, saracco-20}), it still has blue $UVJ$ colors but is expected to move up into the quiescent bin in the next few hundred Myr.
}
\label{fig:UVJ}
\end{figure*}

\section{$UVJ$ Classification and Quiescent Fraction}
\label{sec:passive}

\subsection{Rest-frame Colors and $UVJ$ Classification}
\label{ssec:UVJ} 

The $UVJ$ diagram has become an established method for separating quiescent from star-forming galaxies (\citealt{wuyts-07, williams-09}). Rest-frame $U-V$ and $V-J$ colors were calculated for the 22 spectroscopic members (\S\ref{ssec:specz_membership}) by setting $z=z_{\rm spec}$ and using EAZY (\citealt{brammer-08}). Uncertainties for these colors for each galaxy were calculated by perturbing the flux in each band 10,000 times. The amplitude of the perturbation in each band was determined by sampling from a normal distribution centered on the observed value of flux with width, $\sigma$, equal to the uncertainty in the flux measurement. Each perturbed SED was then fit with EAZY, obtaining a distribution of $U-V$ and $V-J$ rest-frame colors. The 16th and 84th percentile value of that distribution were taken to be the uncertainty in the color for that galaxy.

Figure~\ref{fig:UVJ} shows rest-frame $U-V$ and $V-J$ colors for \pA\ (left) and \pB\ (right). \UMGA\ is shown by a blue star  and \UMGB\ by a green star. The 20 other spectroscopically-confirmed members of \pA\ and \pB\ are shown as solid blue and solid green circles, respectively. The open blue circles show the 24 galaxies classified as photometric members of \pA, and the open green circles show the 12 galaxies classified as photometric members of \pB\ (\S\ref{ssec:photz_membership}).

Also plotted in Figure~\ref{fig:UVJ} is the quiescent selection criteria proposed by \cite{whitaker-11}. 
Notably, \UMGA\ (\pA) is $UVJ$-quiescent, which is consistent with the low levels of star formation estimated from its \Hbeta\ line flux and from UV-to-FIR SED-fitting (\S\ref{ssec:selection}). To our knowledge, there are only two protoclusters above $z>3$ in which the brightest (and most massive) galaxy is both spectroscopically confirmed and $UVJ$-quiescent. These are the SSA22 protocluster at $z=3.09$ (\citealt{Kubo2021}) and \pA~at $z=3.37$.

Other than \UMGA, only one other spectroscopically-confirmed member of protocluster \pA\  has a stellar mass more massive than \logM~$=10.5$. This is galaxy ``181529'' in Table \ref{table:allspeczs} (see also \S\ref{ssec:specz_membership}).
Interestingly, as for \UMGA, galaxy 181529 is $UVJ$ quiescent (solid blue circle in the quiescent wedge in Figure \ref{fig:UVJ}). 

In contrast, the UMG of \pB\ (\UMGB) falls in the post-starburst region of the $UVJ$ color-color diagram. This is consistent with its stellar population determined from NIR spectroscopy (\citealt{marsan-17, saracco-20}). Both \UMGA\ and \UMGB\ appear to have undergone rapid star formation quenching within the last 300 Myr (\citealt{forrest-20b,  saracco-20}). Because \UMGB's star formation quenched so recently and abruptly, it still has blue $UVJ$ colors (Figure~\ref{fig:UVJ}), but it is expected to transition into the quiescent wedge within the next few hundred Myr \citep{marsan-17, merlin-18, belli-19}.

\subsection{Quiescent Fraction of \pA}
\label{ssec:QF}

We calculated the quiescent fraction ($QF$) for \pA\ and also for the field, \ie\ the ratio of the number of quenched galaxies to the total number of galaxies,

\begin{equation} \label{eq:3}
QF = \frac{N^{Q}}{N^{Q} + N^{SF}}.
\end{equation}
 
We calculated the $QF$ only above the DR3 $95\%$ stellar mass completeness limit of \logM~$=10.5$. As described in \S\ref{ssec:photz_membership}, there are 26 members of \pA\ (two spectroscopic and 24 photometric) within a 10 comoving Mpc radius (red circle) of \UMGA, above this stellar mass completeness limit. As noted previously, most of the spectroscopically-confirmed protocluster members are emission-line galaxies which fall below the stellar mass completeness limit and so are not included here in our determination of the $QF$.

We defined the region constituting the ``field'' to be the lower half of the DR3 footprint and the upper part of the right strip i.e. the region below and to the right of the black dashed lines in Figure~\ref{fig:densitymap}. This region was selected because we deemed it least likely to contain any members of the extended \pA/J1000 protocluster system as indicated from the galaxy density contours for the entire COSMOS UltraVISTA field shown in Figure~\ref{fig:densitymap}. Of the 550 galaxies determined photometrically to lie within the redshift slice centered on \UMGA\ (\S\ref{ssec:photz_membership}; shown by the cyan circles in Figure~\ref{fig:densitymap}), a total of 286 lie within these two regions and comprise the field sample. 
The contours in Figure~\ref{fig:UVJ} show the distribution of field galaxies in $UVJ$ color-color space.

Next we classified each galaxy in protocluster \pA\ and in the field either as quiescent or star-forming based on its position in the $UVJ$ diagram (Figure~\ref{fig:UVJ}). As Table~\ref{table:QF} shows, of the 26 members of \pA, 7 are quiescent and 19 star-forming, and of the 286 field galaxies, 47 are quiescent and 239 star-forming.

\begin{figure*}[htbp]
\centering
\includegraphics[width=\textwidth]{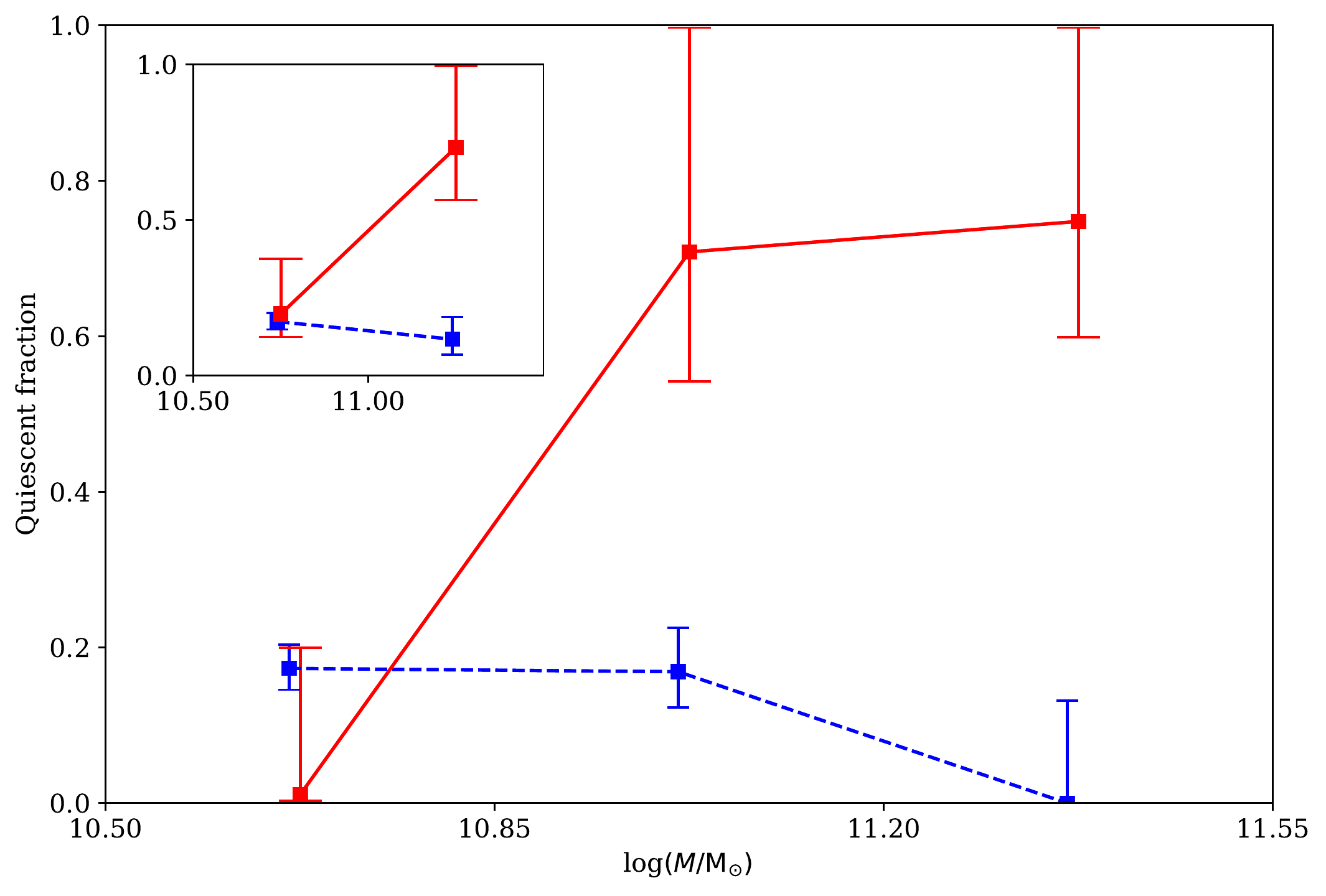}
\caption{Quiescent fraction for protocluster \pA\ (solid red line) and field (dashed blue line) as a function of stellar mass for three mass intervals (two slightly different mass intervals are shown in the inset). The protocluster quiescent fraction shown in the figure is the quiescent fraction in excess of the field \ie\ the field fraction has been subtracted from the ``raw'' protocluster fraction (see \S\ref{ssec:QF} and Table~\ref{table:QF} for details). The quiescent fraction in \pA\ exceeds that of the field, certainly at \logM~$\gtrsim11.0$, and may also be mass dependent,  increasing with stellar mass.
}
\label{fig:QF}
\end{figure*}

The dashed blue lines in Figure~\ref{fig:QF} show the field $QF$ calculated using Equation~\ref{eq:3} as a function of stellar mass for three mass intervals (two slightly different mass intervals are shown in the inset). The values are given in Table~\ref{table:QF}. In order to calculate uncertainties on the field $QF$, we sampled the Poisson distribution for the observed number of quenched galaxies and star-forming galaxies one million times. This was done for each stellar mass bin, enabling us to construct a distribution of quiescent fractions from which we took the $16^{\rm th}$ and $84^{\rm th}$ percentile values to be the uncertainty on the field $QF$. The uncertanties we calculated on the field $QF$ are consistent with those found from analytic approximations (\eg\  \citealt{Gehrels1986}). We also find that our calculated field $QF$ is in good agreement with measurements of the field at $3<z<4$ from the UltraVISTA DR1 catalog (\citealt{Muzzin:2013b}).

In order to calculate the protocluster $QF$, it was necessary to make a correction for contamination by foreground/background field galaxies which had been scattered into the protocluster redshift slice because of redshift uncertainties. The corrected protocluster quiescent fraction was calculated as,
\begin{equation}\label{eq:4}
QF_{corr} = \frac{N_{PC}^{Q} - N_{F}^{Q} (A_{PC}/A_{F})}{N_{PC}^{Q} - N_{F}^{Q} ( A_{PC}/A_{F}) + N^{SF}_{PC} - N^{SF}_{F} ( A_{PC}/A_{F})},
\end{equation}
where $N_{PC}^{Q}$ was the number of quiescent galaxies in the protocluster, $N^{SF}_{F}$ was the number of star-forming galaxies in the field, and $A_{PC}/A_{F}$ was the ratio of the area of the protocluster to the area of the field. The solid red lines in Figure~\ref{fig:QF} show the protocluster $QF$ as a function of stellar mass. 
Uncertainties on the protocluster $QF$ were calculated in a similar manner to those of the field, with the additional step of background subtraction as shown in Equation~\ref{eq:4}. As we did for the field, we sampled the Poisson distribution for the observed number of  quenched and star-forming galaxies in the protocluster. Then the number of quiescent and star-forming galaxies in the field were subtracted from the protocluster, scaled by area. Finally we calculated the fraction of quiescent systems in the resultant protocluster sample. The uncertainty on the protocluster $QF$ were taken to be the $16^{\rm th}$ and $84^{\rm th}$ percentile values of the resulting distribution. 

It is apparent from Figure~\ref{fig:QF} that the $QF$ in \pA\ exceeds that of the field, certainly at \logM~$\gtrsim11.0$, and may also be mass dependent, increasing with increasing stellar mass. We experimented with different $UVJ$ criteria for selecting quiescent and star-forming galaxies, such as the one suggested in \citet{Martis:2016} and also with centering the protocluster on the peak of the density map in Figure~\ref{fig:densitymap} rather than the UMG, but found the $QF$ to be robust to those choices. We also calculated the $QF$ for \pB, but found that protocluster to have a $QF$ much more similar to the field.

\begin{table*}[htbp]
\centering
\caption{\label{table:QF} Quiescent Fractions for Protocluster \pA\ and Field}
\begin{tabular}{cccccc}
\hline
\hline
Stellar Mass & Protocluster $QF$  ($\%$) & Protocluster $QF$  ($\%$) &   Field $QF$ ($\%$) & \# Protocluster Q/SF & \# Field Q/SF\\
\vspace{0.05cm} & (Background corrected) & (Uncorrected) & & &\\
\hline 
\vspace{-0.25cm} & & & & &\\
\vspace{0.05cm} $10.5 \leq$\logM$< 10.85$ & $1.1^{+18.8}_{-1.1}$ & $6.2^{+12.9}_{-5.2}$ & $17.3^{+3.1}_{-2.7}$ & 1/15 & 35/167\\
\vspace{0.05cm} $10.85 \leq$\logM$< 11.2$ & $70.9^{+29.1}_{-16.8}$ & $57.1^{+22.3}_{-24.6}$ & $16.9^{+5.7}_{-4.6}$ & 4/3 & 12/59\\
\vspace{0.1cm} $11.2 \leq$\logM$< 11.55$ & $74.8^{+25.2}_{-15.1}$ & $66.7^{+27.7}_{-41.3}$ & $0.0^{+13.2}_{-0.0}$ & 2/1 & 0/13\\
\hline
\vspace{-0.25cm} & & & & &\\
\vspace{0.05cm} $10.5 \leq$\logM$< 11.0$ & $19.8^{+17.7}_{-7.4}$ & $19.0^{+12.5}_{-8.9}$ & $17.3^{+2.8}_{-2.5}$ & 4/17 & 42/201\\
\vspace{0.1cm} $11.0 \leq$\logM$< 11.55$ & $73.3^{+26.7}_{-16.9}$ & $60.0^{+25.3}_{-30.3}$ & $11.6^{+7.1}_{-4.9}$ & 3/2 & 5/38\\
\hline
\end{tabular}
\end{table*}

 \section{Discussion}
\label{sec:disc}

\subsection{Star Formation and Quenching in Protoclusters}
\label{ssec:diversity}

The existence of such a high $QF$ in \pA\ is in marked contrast to the many known examples of protoclusters which are filled with star-forming galaxies (\citealt{chapman-09, clements-14, dannerbauer-14, casey-15, hung-16, forrest-17, miller-18, cheng-19}).
There are a small number of protoclusters which have previously been reported as having an excess of massive, older or quenched galaxies relative to the field e.g., the $z=2.30$ protocluster in the field of the bright $z = 2.72$ QSO HS 1700+643 (\citealt{Steidel2005, Shapley2005}), the SSA22 protocluster at $z=3.09$ (\citealt{Kubo2021}) and PC 217.96+32.3 at $z= 3.78$ (\citealt{shi-19}).
The discovery of \pA~presented here provides another important counterexample to the suggestion that star-forming or even starbursting galaxies are ubiquitous in protoclusters at $z>2$ (\citealt{casey-16}).  The high $QF$ observed in \pA\ serves to reinforce the viewpoint, instead, that protoclusters exist in a diversity of evolutionary states in the early Universe, and that some systems have quenched remarkably early in the Universe's history.

The high $QF$ in \pA\ raises many interesting questions, not least of which is how the observed quenching has occurred.
A recent analysis of the stellar mass functions (SMFs) of star-forming and quiescent galaxies in a sample of clusters at $1< z< 1.5$ from the GOGREEN survey (\citealt{vanderBurg-20}; see also \citealt{webb-k-20}), returned two surprising results. Firstly, the shapes of the SMFs were identical between cluster and field to high statistical precision, albeit with different normalizations. Secondly, in stark contrast to the mass-independent environmental quenching observed in the local Universe (\citealt{peng-10}),
the observed quenching was strongly dependent on stellar mass. Van der Burg et al. (\citeyear{vanderBurg-20}) concluded that a substantially different quenching mode must operate in dense environments at early times.

Van der Burg et al. (\citeyear{vanderBurg-20}) explored various toy models to interpret their observations showing that they could be reproduced either if the cluster members i) quenched through the same processes as those in the field, but simply did so at an earlier time, a scenario they dubbed ``early mass quenching,'' or ii) underwent a form of environmental quenching which was mass-dependent and, therefore, was significantly different from  mass-independent environmental quenching which has been suggested to occur in the local Universe.

We do not yet know the cause of the quenching observed in \pA\ but it is intriguing to note that it does appear to be mass dependent, which is analogous to the results presented in \citet{vanderBurg-20}. Studies of $1 < z< 1.5$ clusters have determined the environmental quenching timescale, t$_{\rm Q}$, for their members to be about 1 Gyr, meaning that ``classical'' quenching would be expected to begin at $z\sim2$ (\eg\ \citealt{muzzin-14, foltz-18}). While it is certainly possible that environmental processes might be responsible for the quenching observed in \pA\ (scenario ``i'' suggested by \citeauthor{vanderBurg-20}) it seems almost certain that those environmental processes would be different from the processes responsible for ``classical'' quenching  in the low-redshift Universe. Equally possible is that we might be witnessing 
``early mass quenching'' (scenario ``ii'' suggested by \citeauthor{vanderBurg-20}). The fact that we observe a high quenched fraction of galaxies in \pA\ (\S\ref{ssec:QF}) provides additional evidence in support of an ``early mass quenching'' or ``accelerated evolution'' interpretation but additional systems will be required in order to gain further insight.

  \subsection{BCG Formation and the Importance of Self- versus Environmental-Regulation}
\label{ssec:BCG}

The stellar mass of \UMGA\ (\pA) is $\Mstar=2.34^{+0.23}_{-0.34}\times10^{11}~\Mo$, and the stellar mass of  \UMGB\ (\pB)  is 
$\Mstar=2.95^{+0.21}_{-0.20}\times10^{11}~\Mo$. Based on stellar mass, and utilizing the stellar mass-halo mass relation, at $z\sim3$ each of these UMGs would be expected  to reside in a halo of total mass equal to about~$10^{13}~\Mo$ (\citealt{behroozi-13a, behroozi-19, martizzi-12a}). It is, therefore, very likely that these two UMGs are the progenitors of Brightest Cluster Galaxies (BCGs) seen in virialized massive clusters at lower redshift (\citealt{deLucia-07a, vonderLinden-07, bernardi-09, pipino-11, loubser-18, ragone-figueroa-18}).

   Simulations predict that the majority of stars which end up in present-day BCGs form at $z>4$, but that the majority of the present-day stellar mass does not assemble until $z<1.5$, meaning that a BCG would  typically be expected to grow by a factor of about 5-10 between $z=3$ and $z=0$ (\citealt{deLucia-07a,  martizzi-12a, contini-18, ragone-figueroa-18, henden-20}). Our observations are consistent with this scenario whereby BCGs form most of their stars early \ie\ have old stellar ages locally (\citealt{thomas-05, smith-12a, McDermid-15, citro-16, webb-k-20}), with most assembly at late times occurring via mergers/accretion (\citealt{lidman-12a, lidman-13, oliva-altamirano-14a, webb-15b, delahaye-17}). While the exact evolutionary path of \UMGA\ or \UMGB\ cannot be determined from these observations, they are likely destined to evolve to have stellar masses of $\Mstar\gtrsim10^{12}~\Mo$ and be found in Coma-mass (or greater) type halos by the present day.

 This raises another fascinating question --- the importance of self-regulation versus environmental-regulation (\eg\ \citealt{muzzin-12}). Both UMGs have undergone recent, rapid star formation quenching  (\citealt{forrest-20b}), but we are currently unable to determine whether it was the environment itself which was the cause of the quenching, or whether they underwent ``early mass quenching'' {\it because} they happened to lie in an overdense environment (\S\ref{ssec:QF}).

Further analysis will be required to determine whether UMGs are found preferentially in high-density environments or whether \UMGA, which happens to be the oldest UMG in the sample (\citealt{forrest-20b}), is an outlier. Our ongoing MAGAZ3NE survey should yield more information on the environments of these intriguing systems.

 \subsection{The Extended Forming Protocluster \pAB}
\label{ssec:extended}

Protoclusters \pA\ and \pB\ are separated by 17.85 arcmin which corresponds to a comoving separation of 34.7 Mpc or a physical separation of 7.95 Mpc at  $z=3.37$. This separation is  consistent with numerical simulations which follow the early stages of galaxy formation within forming protoclusters over similar scales (\citealt{angulo-12, chiang-13, muldrew-15}). Based on predictions of the size of protoclusters at this epoch, it is quite possible that the  \pAB\  system will evolve into a ``Coma''-type (or even more massive) cluster by the present day. However, because we lack information about the tangential velocities of the members of the extended protocluster \pAB\ system, we cannot conclude this unequivocally. We can only conclude that
we are either witnessing the seeds of a low-redshift massive cluster (if the protoclusters merge) or a super-cluster system (if they do not).

\section{Conclusions}
\label{sec:conclusions}

In this paper we presented the discovery of two new protoclusters, \pA\ (38 members) at  $z=3.37$, and \pB\ (20 members) at $z=3.38$. In contrast to commonly used techniques which have previously been utilized to identify protoclusters, these new systems were discovered using a different approach. They were identified neither by targeting 
LBG, LAE, HAE or DSFG overdensities, nor by targeting commonly utilized ``signposts'' such as HZRG, QSO, or DSFGs. Rather, protoclusters \pA\ and \pB\ were identified
as photometric overdensities around spectroscopically confirmed UMGs
in the course of carrying out the \surveyname\ survey of UMGs and their environments at $3< z<4$. While UMGs may also be thought of as  ``signposts'', it remains to be seen whether they commonly exist in overdense environments and, therefore, whether or not they will prove useful as signposts to identify high-redshift protoclusters.

Notably, and in marked contrast to protoclusters previously reported at this epoch which have been found to predominantly contain star-forming members,  \pA\ was found to have an elevated fraction of quiescent galaxies relative to the coeval field. This high quenched fraction provides a striking and important counterexample to the previously reported pervasiveness of star-forming galaxies in protoclusters at $z > 2$ and suggests, instead, that protoclusters exist in a diversity of evolutionary states in the early Universe.

We do not yet know the cause of the quenching observed in \pA\ but, intriguingly, it appears to be mass dependent (increasing  with  increasing  stellar  mass). This is in stark contrast to mass-independent ``classical'' environmental quenching observed in the local Universe, but in agreement with recent results at  $1 < z< 1.5$ from the GOGREEN survey. Whether we are witnessing ``early mass quenching'' or non-traditional ``environmental quenching'' will require larger samples to determine.

 Both UMGs have undergone recent, rapid star formation quenching (\citealt{forrest-20b}), but we are currently unable to determine whether it was the environment itself which was the cause of the quenching, or whether they underwent ``early mass quenching'' {\it because} they happened to lie in an overdense environment. Determining the relative importance of self-regulation versus environmental-regulation will also require larger samples.

Based on their stellar mass and the stellar mass--halo mass relation, we concluded that \UMGA\ and \UMGB\ reside in halos of  $\sim10^{13}~\Mo$ at $z\sim3$. They may well be the descendants of the population of highly dust-obscured massive star-forming galaxies discovered at $z > 5$ (see \citealt{forrest-20a, forrest-20b} for more discussion). It is also highly likely that \UMGA\ and \UMGB\ are the progenitors of BCGs seen in virialized massive clusters at lower redshift. While the exact evolutionary paths of \UMGA\ and \UMGB\ cannot be predicted, the observations presented here show that very massive galaxies can form and quench surprisingly early during protocluster formation.  

  Protoclusters \pA\ and \pB\ are separated by 35 comoving Mpc, in good agreement with predictions from simulations regarding the size of ``Coma''-type cluster progenitors  forming at this epoch. Irrespective of whether or not \pA\ and \pB\ will actually merge by $z=0$, we are undoubtedly witnessing the seeds of a low-redshift super-cluster system.

  The \pA/J1000 protocluster system presented here was discovered in a field totaling only about 0.84 deg$^{2}$ in area. Future ground and space telescopes with the capability to survey significantly wider areas, \eg\  the {\it James Webb Space Telescope} (\citealt{gardner-06}), the {\it Vera C. Rubin Observatory} (\citealt{LSSTv2-09}), the {\it Euclid Space Telescope} (\citealt{laureijs-11}), and the {\it Nancy Grace Roman Space Telescope} (\citealt{spergel-15}) will undoubtedly facilitate the discovery of larger samples, allowing better insight into the uniqueness of \pA,  and helping to propel our understanding of the formation of UMGs and protoclusters into the even earlier Universe.

\section*{Acknowledgments}

This work was supported by the National Science Foundation through grants AST-1513473, AST-1517863, AST-1518257, and AST-1815475, by HST program number GO-15294, by UNAB internal grant DI-12-19/R, and by grant numbers 80NSSC17K0019 and NNX16AN49G issued through the NASA Astrophysics Data Analysis Program (ADAP). Support for program number GO-15294 was provided by NASA through a grant from the Space Telescope Science Institute, which is operated by the Association of Universities for Research in Astronomy, Incorporated, under NASA contract NAS5-26555.

The data presented herein were obtained at the W. M. Keck Observatory, which is operated as a scientific partnership among the California Institute of Technology, the University of California and the National Aeronautics and Space Administration. The Observatory was made possible by the generous financial support of the W. M. Keck Foundation. The authors wish to recognize and acknowledge the very significant cultural role and reverence that the summit of Maunakea has always had within the indigenous Hawaiian community.  We are most fortunate to have the opportunity to conduct observations from this mountain.

\bibliographystyle{aasjournal}
\bibliography{mergedbib}

\begin{figure*}[htp]
\centering
\begin{tabular}{c}
\includegraphics[width=\textwidth]{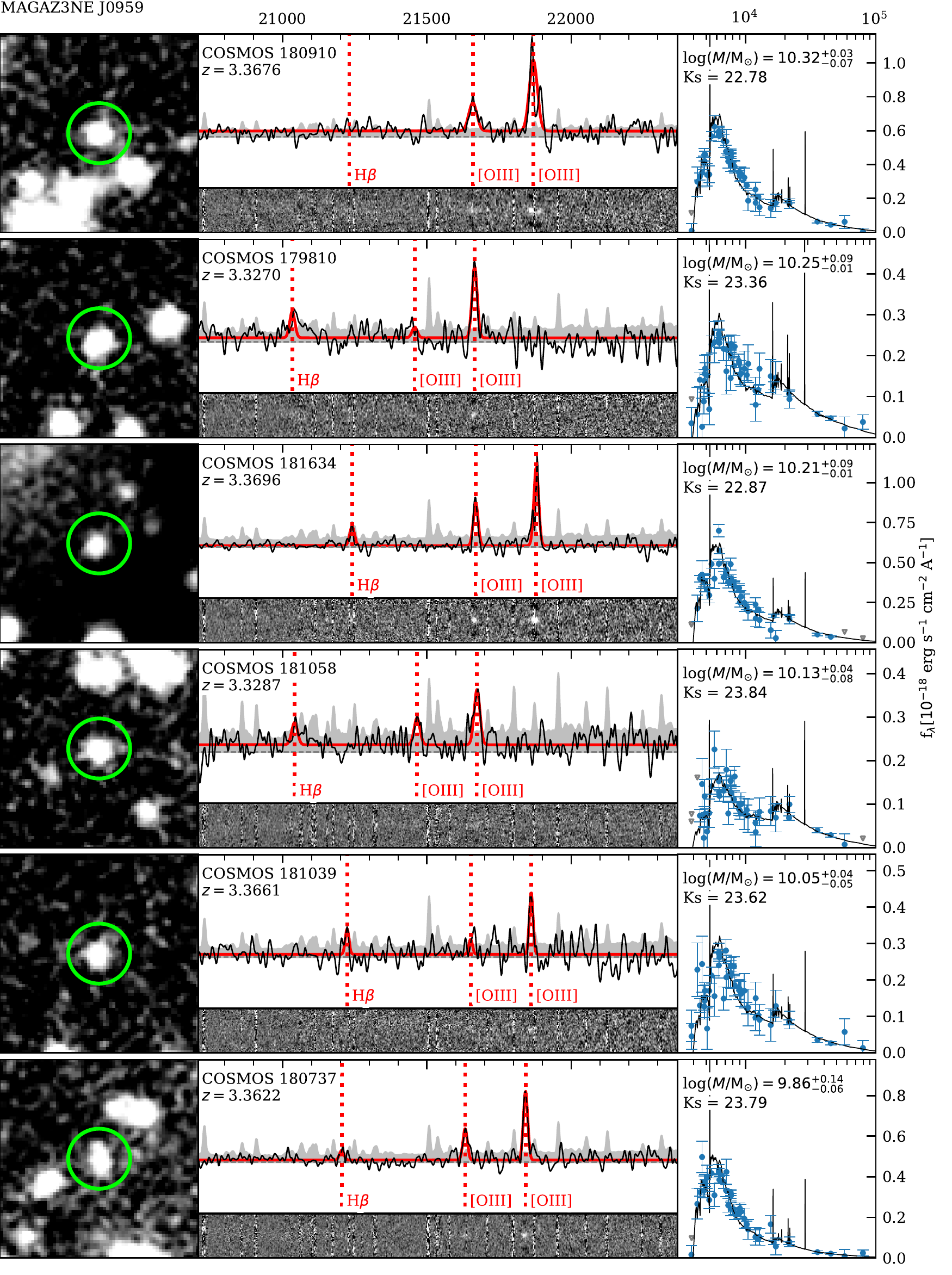}
\end{tabular}

\end{figure*}
\begin{figure*}[htp]
\centering
\begin{tabular}{c}
\includegraphics[width=\textwidth]{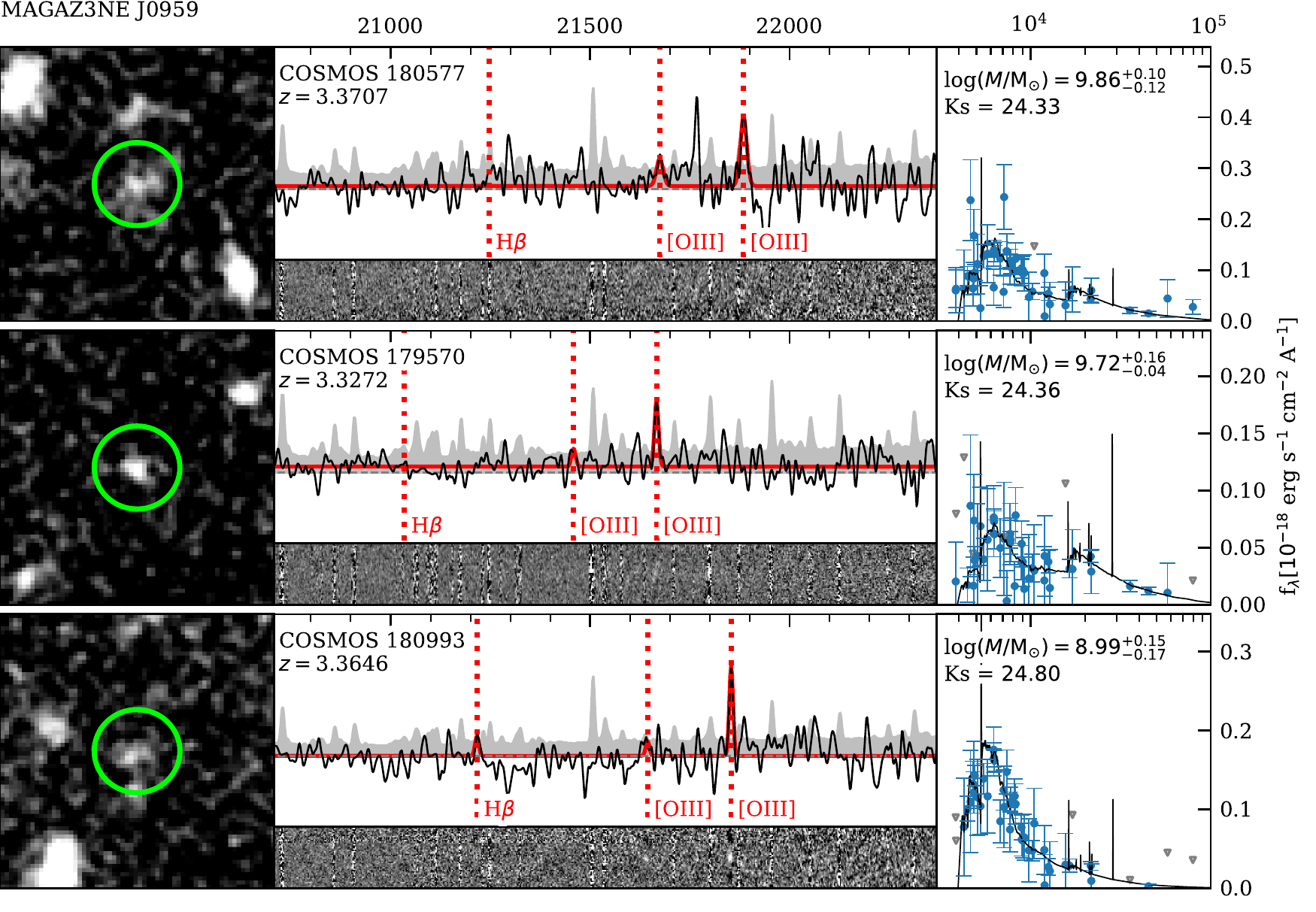}\\
\includegraphics[width=\textwidth]{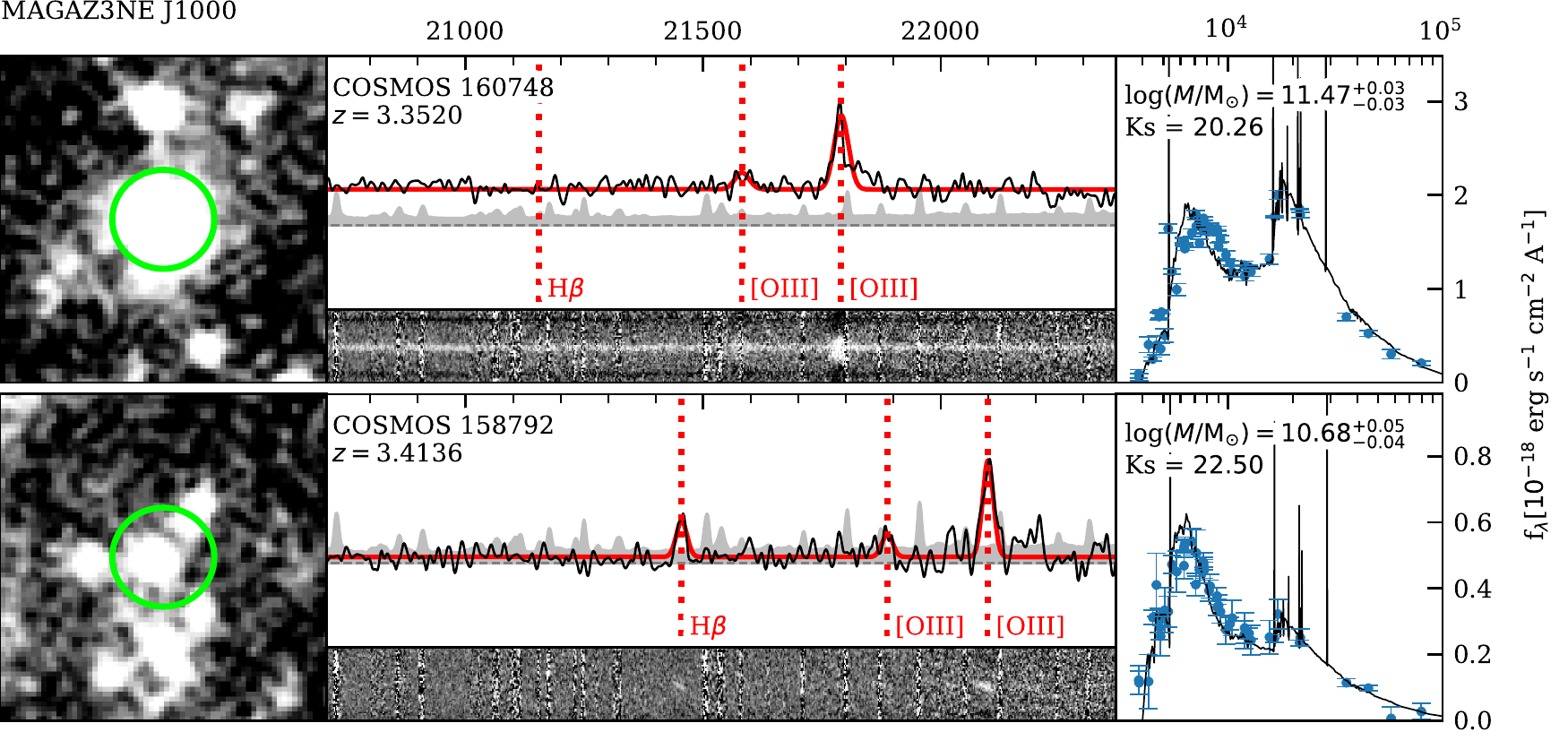}
\end{tabular}
\end{figure*}

\begin{figure*}[htp]
\centering
\begin{tabular}{c}
\includegraphics[width=\textwidth]{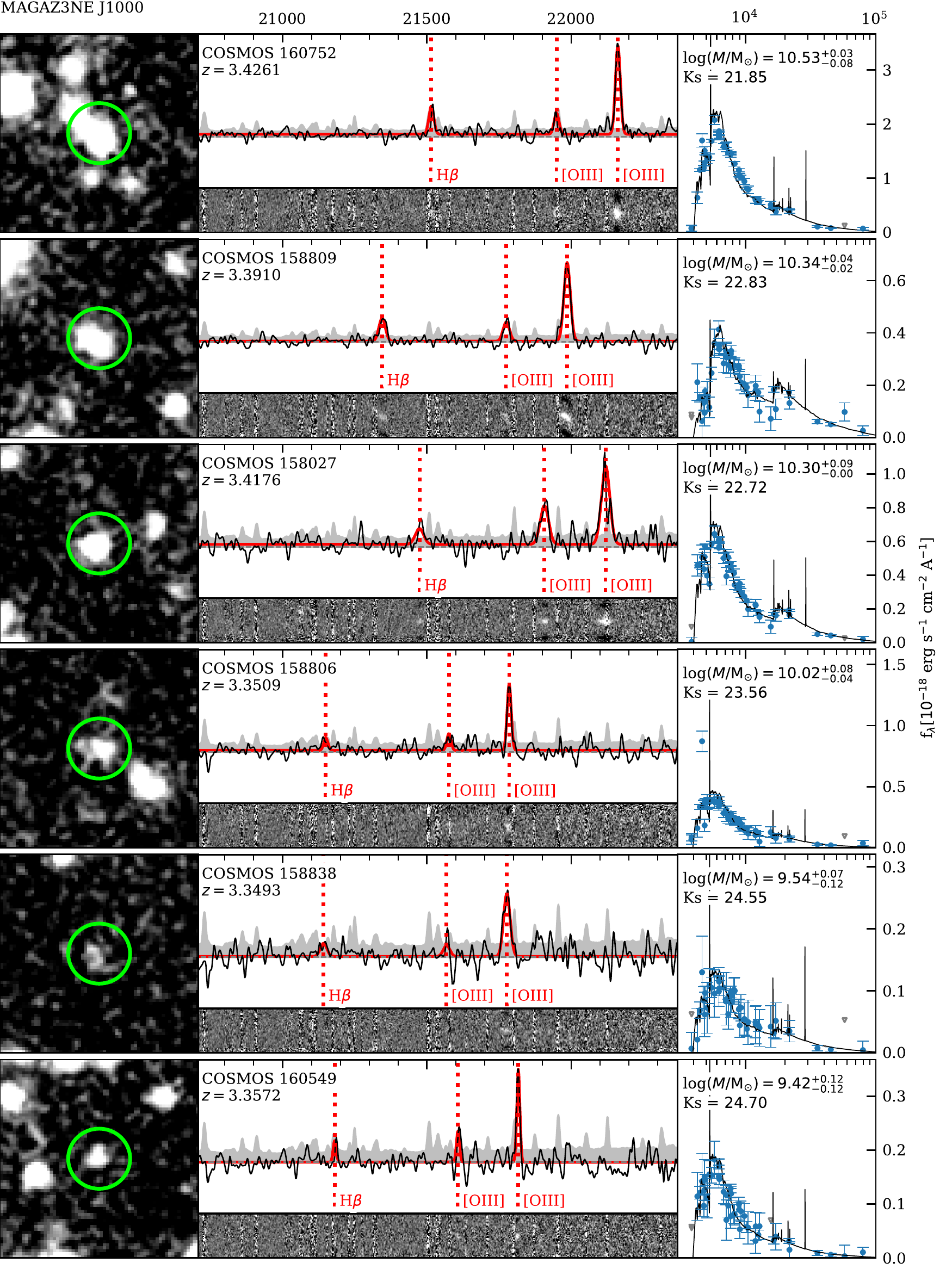}
\end{tabular}
\end{figure*}

\clearpage

\end{document}